\documentclass[11pt]{article}

\usepackage{indentfirst}
\usepackage{amsmath}
\usepackage{amssymb}
\usepackage{amsfonts}
\usepackage[dvipsnames]{xcolor}
\usepackage[pdftex]{graphicx}
\usepackage{subfigure}
\usepackage{subcaption}
\usepackage{float}
\usepackage{multirow}
\usepackage{makecell}

\usepackage{geometry}
\usepackage[backend=bibtex, style=numeric-comp, sorting=none, maxnames=99]{biblatex}
\usepackage{hyperref}
\hypersetup
{
    colorlinks = true,
    linkcolor = red,
    citecolor = blue,
    urlcolor = PineGreen
}

\graphicspath{{./Figs/}}

\begin{document}
\begin{titlepage} \setcounter{page}{0}
\begin{center}
    \vspace*{1.0cm}
    {\Large\bf Phase Transitions with Lyapunov Exponents under Einstein and String Frames in Dilatonic Reissner--Nordstr\"om--AdS Black Holes}
    \\ \vspace{2.0cm}
    {$\mbox{Hocheol Lee}$}\footnote{\it email: insaying@dongguk.edu}, \quad
    {$\mbox{Bogeun Gwak}$}\footnote{\it email: rasenis@dgu.ac.kr}
    \\ \vspace{0.2cm}
    {\small \it Department of Physics, Dongguk University, Seoul 04620, Republic of Korea}
    \\ \vspace{2.0cm}
\end{center}

\begin{center}
\begin{abstract}
    We investigate Lyapunov exponents as dynamical probes of black hole phase transitions in dilatonic Reissner--Nordstr\"om--AdS black holes within Einstein--Maxwell--dilaton theory. The thermodynamic quantities and the Lyapunov exponent of charged probe particles were analyzed in both the Einstein and string frames, thus providing a direct comparison between the thermodynamic phase structure of the black hole and that captured by the Lyapunov exponent. Thermodynamic quantities, including the Hawking temperature and Wald entropy, remained constant under conformal frame transformations, yielding identical phase structures in the two frames. In contrast, the Lyapunov exponent exhibited non-trivial frame dependence for massive probe particles due to dilaton coupling, while no frame dependence was found in the massless limit. Numerical analysis revealed that the phase structure features captured by the Lyapunov exponent, including characteristic cusp behavior and transition points, were independent of the choice of frame, despite the Lyapunov exponent itself being frame-dependent. Therefore, the Lyapunov exponent exhibited frame-dependent values, while the critical structure it captures remained constant across conformal frames.
\end{abstract}
\end{center}
\end{titlepage}

\clearpage
{
    \renewcommand{\baselinestretch}{1.4}
    \normalsize
    \setlength{\parskip}{7pt}
    \tableofcontents
}
\clearpage

\newpage
\section{Introduction} \label{sec:introduction}
    Black holes are of prime importance in modern theoretical physics, which encompasses general relativity, quantum mechanics, and thermodynamics. The discoveries of black hole entropy~\cite{Bekenstein:1973ur} and Hawking radiation~\cite{Hawking:1975vcx} have established black holes as thermodynamic systems characterized by entropy and temperature. The thermodynamic study of black holes has revealed phase transitions and critical phenomena analogous to those encountered in ordinary thermodynamic systems. These phenomena are particularly rich in asymptotically anti-de Sitter (AdS) spacetimes. The Hawking--Page transition~\cite{Hawking:1982dh} between thermal AdS space and a stable black hole phase demonstrated the existence of phase transitions in gravitational systems. Charged AdS black holes exhibit even richer thermodynamic behavior, including first-order phase transitions between small and large black hole phases and critical phenomena analogous to those found in a van der Waals fluid~\cite{Chamblin:1999tk}. These developments have established black holes as theoretical laboratories for studying thermodynamic phase structures and critical phenomena in gravitational systems.
    
    Extensions of Einstein gravity often introduce additional scalar fields, generating a broader class of black hole solutions with non-trivial geometric and thermodynamic properties. Among these, dilatonic black holes form an important class of solutions in Einstein--Maxwell--dilaton theory, which has emerged as a low-energy effective description of heterotic string theory~\cite{Gibbons:1987ps, Garfinkle:1990qj}. The non-minimal coupling of the dilaton field to the gravitational and electromagnetic sectors introduces an additional scalar degree of freedom, thereby altering the geometric and thermodynamic properties of black holes. Accordingly, the inclusion of the dilaton field can alter the thermodynamic phase structure of charged black holes in asymptotically AdS spacetimes. A distinctive feature of Einstein--Maxwell--dilaton theory is the existence of multiple conformal frames related to each other by conformal transformations. This theory admits both the Einstein frame, in which the gravitational sector takes the standard Einstein--Hilbert form, and the string frame, in which the dilaton field is non-minimally coupled to spacetime curvature. Although the two frames are related by a conformal transformation and are generally considered equivalent descriptions of a given black hole solution, geometric and physical quantities can differ depending on the choice of frame. Whether a given quantity remains constant irrespective of the choice of frame or exhibits explicit frame dependence has been extensively studied~\cite{Magnano:1993bd, Faraoni:1998qx} and has continued to attract research attention in recent years~\cite{Zhang:2025aak, Valadao:2025kef, Alho:2026qxz}. Thus, owing to the coexistence of the Einstein and string frame descriptions, dilatonic black holes are well suited for investigating the frame dependence of thermodynamic and dynamical quantities under conformal frame transformations.

    Beyond thermodynamic quantities, dynamical probes provide a complementary perspective on the physical properties of black holes. The Lyapunov exponent has emerged as one of the more widely studied dynamical probes of black hole spacetimes~\cite{Hashimoto:2016dfz, Zhao:2018wkl, Lei:2020clg, Kan:2021blg, Lei:2021koj, Gwak:2022xje, Yu:2022tlr, Gao:2022ybw, Chen:2022tbb, Hashimoto:2022kfv, Chakrabortty:2022kvq, Jeong:2023hom, Chen:2023wph, Yu:2023spr, Li:2023bgn, Lei:2023jqv, Prihadi:2023tvr, Prihadi:2023qmk, Xie:2023tjc, Dutta:2023yhx, Karan:2023hfk, Park:2023lfc, Kumara:2024obd, Giataganas:2024hil, Lei:2024qpu, Das:2024iuf, Singh:2024qfw, Dutta:2024rta, Gallo:2024wju, Ciou:2025ygb, R:2025gok, Lee:2025vih, An:2025xmb, Targema:2025opv, Li:2026ocp, Yang:2026gcv, Dalui:2026msw, Li:2026ebc, Targema:2026anu}. Defined as the local exponential growth rate of perturbations in unstable particle trajectories, the Lyapunov exponent carries information about the effective potential that drives particle motion and about the underlying spacetime geometry. Recent studies have revealed that the Lyapunov exponent can capture the thermodynamic phase structure of black holes~\cite{Guo:2022kio, Yang:2023hci, Lyu:2023sih, Du:2024uhd, Shukla:2024tkw, Gogoi:2024akv, Awal:2025irl, Yang:2025fvm, Kumar:2025kzt, Guo:2025pit, Bezboruah:2025udi, Zhang:2025cdx, Xie:2025auj, Gao:2025rep, Becar:2026epq, Cheng:2026dnd, Yovkochev:2026wfg, Awal:2026diz}. In various AdS black hole systems, multi-valued branches and critical phenomena associated with the Lyapunov exponent mirror thermodynamic phase transitions, indicating that dynamical probes can capture features that are traditionally extracted from thermodynamic quantities. Conformal frame dependence naturally extends to the relation between dynamical and thermodynamic probes. A previous study revealed that the Lyapunov exponent of charged particles in dilatonic black hole spacetimes is explicitly dependent on the choice of frame~\cite{Lee:2025ias}. In contrast, thermodynamic quantities are conventionally considered frame-invariant, in line with the general expectation that physical observables are independent of arbitrary frame choices. The contrast between the frame dependence of the Lyapunov exponent and the independence of thermodynamic quantities from the choice of frame raises the question of whether the phase structure captured by the Lyapunov exponent remains consistent with the thermodynamic phase structure under conformal frame transformations. This is also related to whether the Lyapunov exponent identifies the same transition points and critical behavior in different frames, or whether its frame-dependent value modifies the phase transition features it captures. Analyzing this problem provides a direct comparison between thermodynamic and dynamical probes and further elucidates the physical implications of frame dependence in dilatonic black hole spacetimes.

    In this study, the frame dependence of thermodynamic and dynamical probes in dilatonic Reissner--Nordstr\"om--AdS black holes was investigated. Both the Einstein and string frames were considered, and thermodynamic properties and particle dynamics in the two conformal frames were analyzed in parallel. The analysis focused on the Lyapunov exponent associated with the motion of a charged probe particle, whose dynamics was directly influenced by coupling to the dilaton field. The relationship between dynamical probes and black hole thermodynamics under conformal frame transformations was then examined. The thermodynamic phase structure was compared with the behavior of the Lyapunov exponent in the Einstein and string frames to determine whether the phase transition features captured by the Lyapunov exponent remained consistent with the thermodynamic phase structure in different conformally related descriptions.

    The remainder of this paper is organized as follows. Section~\ref{review:DRN-AdS} reviews the properties of the dilatonic Reissner--Nordstr\"om--AdS black hole solution based on the Einstein--Maxwell--dilaton theory. Section~\ref{sec:frame_dependence} presents the derivations of the thermodynamic quantities and Lyapunov exponent in the Einstein and string frames. Section~\ref{sec:numerical_results} provides the numerical results and analyzes the frame dependence of the thermodynamic quantities and Lyapunov exponent in both asymptotically flat and asymptotically AdS spacetimes, including an examination of the Lyapunov exponent as a probe of thermodynamic phase structures for different conformal frame transformations. Section~\ref{sec:conclusion} concludes the paper with a summary of the results and a discussion of future research directions.

\section{Review of the Dilatonic Reissner--Nordstr\"om--AdS Black Holes} \label{review:DRN-AdS}
    This section reviews the properties of dilatonic Reissner--Nordstr\"om--AdS (RN--AdS) black holes and the Lyapunov exponents for particle motion. In Einstein--Maxwell--dilaton (EMD) theory, the four-dimensional low-energy effective action in the Einstein frame, with $G = c = 1$, takes the following form:
\begin{equation}
    I_\mathrm{E} = \frac{1}{16 \pi} \int \mathrm{d}^4x \sqrt{-g_\mathrm{E}} \left( R_\mathrm{E} - \frac{1}{2} g_\mathrm{E}^{\mu \nu} \partial_\mu \phi \partial_\nu \phi - e^{-\alpha \phi} g_\mathrm{E}^{\mu \rho} g_\mathrm{E}^{\nu \sigma} F_{\mu \nu} F_{\rho \sigma} - U(\phi, \Lambda) \right),
\end{equation}
    where the subscript $\mathrm{E}$ denotes quantities in the Einstein frame; $R$ is the Ricci scalar; $\phi$ is the dilaton field; $F_{\mu \nu}$ is the electromagnetic tensor, which is non-minimally coupled to the dilaton field with coupling parameter $\alpha$; and $U(\phi, \Lambda)$ is the dilaton potential including the cosmological constant $\Lambda$~\cite{Gao:2005xv}, which is defined as follows:
\begin{equation}
    U(\phi, \Lambda) = \frac{2 \Lambda}{3 \left( 1 + \alpha^2 \right)^2} \left[ \left( 3 - \alpha^2 \right) e^{\alpha \left( \phi - \phi_\infty \right)} + \alpha^2 \left( 3 \alpha^2 - 1 \right) e^{-\frac{\phi - \phi_\infty}{\alpha}} + 8 \alpha^2 e^{-\frac{( 1 - \alpha^2 ) \left( \phi - \phi_\infty \right)}{2 \alpha}} \right].
\end{equation}
    The electromagnetic potential and electromagnetic tensor are given by
\begin{equation}
    \mathbf{A} = A_t(r)\mathrm{d}t, \qquad F_{\mu \nu} = \partial_\mu A_\nu - \partial_\nu A_\mu.
\end{equation}
    The field equations for the metric, dilaton, and gauge field are given by
\begin{align}
    0 &= R^\mathrm{E}_{\mu\nu} - \frac{1}{2} \partial_\mu \phi \partial_\nu \phi + \frac{1}{2} e^{-\alpha \phi} g_\mathrm{E}^{\rho \gamma} g_\mathrm{E}^{\sigma \tau} F_{\gamma \tau} F_{\rho \sigma} g^\mathrm{E}_{\mu\nu} + 2 e^{-\alpha \phi} g_\mathrm{E}^{\rho \sigma} F_{\mu \rho} F_{\sigma \nu} - \frac{1}{2} U g^\mathrm{E}_{\mu\nu},
    \\
    0 &= g_\mathrm{E}^{\rho \sigma} \nabla^\mathrm{E}_\rho \nabla^\mathrm{E}_\sigma \phi + \alpha e^{-\alpha \phi} g_\mathrm{E}^{\mu \rho} g_\mathrm{E}^{\nu \sigma} F_{\mu \nu} F_{\rho \sigma} - \frac{\partial U}{\partial \phi},
    \\
    0 &= \nabla^\mathrm{E}_\rho \left( e^{-\alpha \phi} g_\mathrm{E}^{\rho \sigma} F_{\sigma \mu} \right).
\end{align}
    For a general static and spherically symmetric metric in the Einstein frame,
\begin{equation}
    \mathrm{d}s_\mathrm{E}^2 = - f(r) \mathrm{d}t^2 + \frac{\mathrm{d}r^2}{g(r)} + \mathcal{R}(r)^2 \left( \mathrm{d}\theta^2 + \sin^2\theta \mathrm{d}\varphi^2 \right),
    \label{eq:metric_Einstein}
\end{equation}
    the solution for the dilatonic RN--AdS black hole is expressed as follows:
\begin{align}
    f(r) &= g(r) = \left( 1 - \frac{r_+}{r} \right) \left( 1 - \frac{r_-}{r} \right)^\frac{1 - \alpha^2}{1 + \alpha^2} - \frac{\Lambda}{3} r^2 \left( 1 - \frac{r_-}{r} \right)^\frac{2 \alpha^2}{1 + \alpha^2},
    \label{eq:f}
    \\
    \mathcal{R}(r) &=  r \left( 1 - \frac{r_-}{r} \right)^\frac{\alpha^2}{1 + \alpha^2},
    \\
    \phi(r) &= \phi_\infty + \frac{2 \alpha}{1 + \alpha^2} \log\left( 1 - \frac{r_-}{r} \right),
    \\
    A_t(r) &= - \frac{e^{\alpha \phi_\infty} Q_\mathrm{E}}{r}, \qquad Q_\mathrm{E}^2 = \frac{e^{-\alpha \phi_\infty} r_+ r_-}{1 + \alpha^2},
    \label{eq:black_hole_charge}
\end{align}
    where $\phi_\infty$ represents the asymptotic value of the dilaton field at spatial infinity, and $Q_\mathrm{E}$ denotes the black hole charge, obtained from a flux integral over the sphere at spatial infinity
\begin{equation}
    Q_\mathrm{E} = \frac{1}{4 \pi} \oint_{S^2_\infty} \sqrt{-g_\mathrm{E}} \, e^{-\alpha \phi} g_\mathrm{E}^{t \rho} g_\mathrm{E}^{r \sigma} F_{\rho \sigma} \, \mathrm{d}\theta \mathrm{d}\varphi.
\end{equation}
    For $\alpha = 0$ and $\alpha = 1$, the solution reduces to the electrically charged RN--AdS and Gibbons--Maeda--Garfinkle--Horowitz--Strominger black holes~\cite{Gibbons:1987ps, Garfinkle:1990qj}, respectively. In asymptotically flat spacetime $( \Lambda = 0 )$, $r_+$ and $r_-$ correspond to the outer event horizon $r_\mathrm{h}$ and the possible inner Cauchy horizon, respectively, with the latter present for $\alpha^2 < 1$. In asymptotically AdS spacetime $( \Lambda < 0 )$, $r_+$ does not represent the event horizon, which is instead located at $r_\mathrm{h}$ with $r_\mathrm{h} < r_+$.

    The thermodynamic properties of the dilatonic RN--AdS black hole are now considered. The Hawking temperature in the Einstein frame is given by
\begin{equation}
    T_\mathrm{E} = \frac{1}{4 \pi r_\mathrm{h}^2}\left[ 3 r_+ - 2 r_\mathrm{h} - r_- - \frac{4\left( r_+ - r_\mathrm{h} \right) r_-}{\left( 1 + \alpha^2 \right) r_\mathrm{h}} \right] \left( 1 - \frac{r_-}{r_\mathrm{h}} \right)^{- \frac{2 \alpha^2}{1 + \alpha^2}},
\end{equation}  
    where asymptotically flat spacetime is recovered in the limit $r_+ = r_\mathrm{h}$. The black hole entropy is expressed as follows:
\begin{equation}
    S_\mathrm{E} = \pi r_\mathrm{h}^2 \left( 1 - \frac{r_-}{r_\mathrm{h}} \right)^\frac{2 \alpha^2}{1 + \alpha^2}.
\end{equation}
    The conserved mass is defined by the Brown--York quasi-local formalism~\cite{Brown:1992br}. In the present case, where the spacetime is not exactly asymptotically AdS, a counterterm is required to ensure a finite charge, leading to the modified Brown--York quasilocal mass~\cite{Sheykhi:2009pf, Bhattacharya:2023ycc}, given by
\begin{equation}
    M_\mathrm{E} = \frac{1}{2} \left( r_+ + \frac{1 - \alpha^2}{1 + \alpha^2} r_- \right).
\end{equation}
    The first law of black hole thermodynamics is as follows:
\begin{equation}
    \mathrm{d}M_\mathrm{E} = T_\mathrm{E} \mathrm{d}S_\mathrm{E} + \Phi_\mathrm{E} \mathrm{d}Q_\mathrm{E},
\end{equation}
    where the quantity $\Phi$ represents the electric potential, defined by
\begin{equation}
    \Phi_\mathrm{E} = \left. \xi_{(t)}^\mu A_\mu \right|_{r \to \infty} - \left. \xi_{(t)}^\mu A_\mu \right|_{r = r_\mathrm{h}} = \frac{e^{\alpha \phi_\infty} Q_\mathrm{E}}{r_\mathrm{h}},
\end{equation}
    with the timelike Killing vector $\boldsymbol{\xi}_{(t)} = (1, \, 0, \, 0, \, 0)$.

\section{Frame Dependence of the Thermodynamic Quantities and Lyapunov Exponents} \label{sec:frame_dependence}
    The frame dependence of the thermodynamic and dynamical quantities in static, spherically symmetric spacetimes is investigated. The conformal transformation induced by the dilaton field modifies the geometric representation of a given spacetime and requires a comparison between the Einstein and string frames. The quantities under consideration include the Hawking temperature, black hole entropy, and Lyapunov exponent, which characterizes orbital instability in particle motion. This analysis is conducted to elucidate the effects of non-minimal dilaton coupling by adopting the dilatonic RN--AdS black hole as a representative background for a comparative study of frame dependence.

\subsection{Thermodynamic Quantities}
    The most general static, spherically symmetric metric in the Einstein frame~\eqref{eq:metric_Einstein}, assuming that the metric function $f(r)$ vanishes at the event horizon $r_\mathrm{h}$ and is asymptotically flat at spatial infinity, is given by
\begin{equation}
    f(r_\mathrm{h}) = 0, \qquad \lim_{r \to \infty} f(r) = 1.
\end{equation}    
    Because the spacetime is static, it admits a timelike Killing vector
\begin{equation}
    \xi_\mathrm{E}^\mu = (1, \, 0, \, 0, \, 0).
    \label{eq:Killing_vector_Einstein}
\end{equation}
    The norm of the Killing vector in the Einstein frame~\eqref{eq:Killing_vector_Einstein} is given by
\begin{equation}
    \xi_\mathrm{E}^2 = -f(r),
\end{equation}
    which is normalized at spatial infinity as follows:
\begin{equation}
    \lim_{r \to \infty} \xi_\mathrm{E}^2 = -1.
\end{equation}
    The Hawking temperature $T$ in the Einstein frame is obtained from the surface gravity $\kappa$ as
\begin{equation}
    T_\mathrm{E} = \frac{\kappa_\mathrm{E}}{2 \pi} = \frac{1}{2 \pi} \sqrt{-\frac{1}{2} \left( \nabla_\mathrm{E} \xi_\mathrm{E} \right)^2} = \frac{1}{4 \pi} \sqrt{\frac{g(r_\mathrm{h})}{f(r_\mathrm{h})}} f'(r_\mathrm{h}),
    \label{eq:Hawking_temperature}
\end{equation}
    in which a prime symbol denotes differentiation with respect to $r$. For the corresponding spacetime in the string frame,
\begin{equation}
    \mathrm{d}s_\mathrm{S}^2 \equiv e^\phi \mathrm{d}s_\mathrm{E}^2 = e^\phi(r) \left[ - f(r) \mathrm{d}t^2 + \frac{\mathrm{d}r^2}{g(r)} + \mathcal{R}^2(r) \left( \mathrm{d}\theta^2 + \sin^2\theta \mathrm{d}\varphi^2 \right) \right].
    \label{eq:metric_string}
\end{equation}
    Under the conformal transformation, the canonical normalization of the Killing vector in the string frame~\eqref{eq:metric_string} at infinity requires
\begin{equation}
    \xi_\mathrm{S}^\mu = (e^{-\frac{\phi_\infty}{2}}, \, 0, \, 0, \, 0),
\end{equation}
    where $\phi_\infty$ denotes the asymptotic value of the dilaton field. The norm of the Killing vector in the string frame is given by
\begin{equation}
    \xi_\mathrm{S}^2 = - e^{\phi - \phi_\infty} f(r), \qquad \lim_{r \to \infty} \xi_\mathrm{S}^2 = -1,
\end{equation}
    which ensures consistency with the normalization in the Einstein frame. The Hawking temperature in the string frame is as follows:
\begin{equation}
    T_\mathrm{S} = \frac{\kappa_\mathrm{S}}{2 \pi} = \left. \frac{e^{-\frac{\phi_\infty}{2}}}{4 \pi} \sqrt{\frac{g(r)}{f(r)}} f'(r) \right|_{r = r_\mathrm{h}},
\end{equation}
    which yields
\begin{equation}
    T_\mathrm{S} = e^{-\frac{\phi_\infty}{2}} T_\mathrm{E}.
    \label{eq:temperature_Einstein_string}
\end{equation}
    The relationship~\eqref{eq:temperature_Einstein_string} shows that the Hawking temperatures differ by a constant factor depending on the asymptotic value of the dilaton field. The mass and electric potential are analogously dependent on $\phi_\infty$, whereas the electric charge remains constant,
\begin{equation}
    M_\mathrm{S} = e^{-\frac{\phi_\infty}{2}} M_\mathrm{E}, \quad \Phi_\mathrm{S} = e^{-\frac{\phi_\infty}{2}} \Phi_\mathrm{E}, \quad Q_\mathrm{S} = Q_\mathrm{E}.
\end{equation}
    Because the asymptotic dilaton value can be redefined by a constant shift $\phi \to \phi + \phi_\infty$, this shift corresponds to a redefinition of the asymptotic value of the dilaton field. Without loss of generality, the asymptotic value can thus be chosen as $\phi_\infty = 0$, for which the Hawking temperature, mass, and electric potential coincide with the corresponding values in the Einstein and string frames. For asymptotically AdS spacetime, the norm of the timelike Killing vector diverges at spatial infinity. In the context of the AdS/conformal field theory correspondence, the canonical normalization of the Killing vector is defined in terms of the conformal boundary metric. Thus, the expression for the Hawking temperature is applicable to asymptotically AdS spacetime without modification, yielding identical Hawking temperatures in the Einstein and string frames for the dilatonic RN--AdS black hole
\begin{equation}
    T_\mathrm{H} = \frac{1}{4 \pi r_\mathrm{h}^2}\left[ 3 r_+ - 2 r_\mathrm{h} - r_- - \frac{4\left( r_+ - r_\mathrm{h} \right) r_-}{\left( 1 + \alpha^2 \right) r_\mathrm{h}} \right] \left( 1 - \frac{r_-}{r_\mathrm{h}} \right)^{- \frac{2 \alpha^2}{1 + \alpha^2}}.
\end{equation}

    The black hole entropy in the Einstein frame is now considered. The Bekenstein--Hawking entropy~\cite{Bekenstein:1973ur, Hawking:1975vcx} is obtained using the horizon area $\mathcal{A}$ as follows:
\begin{equation}
    S_\mathrm{E} = \frac{\mathcal{A}_\mathrm{E}}{4} = \frac{1}{4} \int_0^{2 \pi} \mathrm{d}\varphi \int_0^\pi \mathcal{R}(r_\mathrm{h})^2 \sin\theta \, \mathrm{d}\theta = \pi \mathcal{R}(r_\mathrm{h})^2.
    \label{eq:entropy_Einstein}
\end{equation}
    Following the computation~\eqref{eq:entropy_Einstein}, the naive Bekenstein--Hawking entropy in the string frame is given by
\begin{equation}
    \frac{\mathcal{A}_\mathrm{S}}{4} = \frac{e^{\phi(r_\mathrm{h})}}{4} \int_0^{2 \pi} \mathrm{d}\varphi \int_0^\pi \mathcal{R}(r_\mathrm{h})^2 \sin\theta \, \mathrm{d}\theta = \pi e^{\phi(r_\mathrm{h})} \mathcal{R}(r_\mathrm{h})^2 = e^{\phi(r_\mathrm{h})} S_\mathrm{E},
\end{equation}
    which differs from that in the Einstein frame. Thus, the Bekenstein--Hawking entropy shows frame dependence due to the non-vanishing dilaton field at the horizon, suggesting that the horizon area is insufficient to capture the full gravitational entropy in the presence of non-minimal coupling to the dilaton field. However, because the Bekenstein--Hawking entropy in the Einstein frame $S_\mathrm{E}$ coincides with the Wald entropy~\cite{Wald:1993nt}, the Wald entropy formalism should be employed directly in the string frame for consistency. Thus, the low-energy effective action of EMD theory is considered in the string frame
\begin{equation}
    I_\mathrm{S}  = \frac{1}{16 \pi} \int \mathrm{d}^4x \sqrt{-g_\mathrm{S}} \, \mathcal{L}_\mathrm{S} = \frac{1}{16 \pi} \int \mathrm{d}^4x \sqrt{-g_\mathrm{S}} \, e^{-\phi} \left( R_\mathrm{S} + g_\mathrm{S}^{\mu \nu} \partial_\mu \phi \partial_\nu \phi - e^{\left( 1 -\alpha \right) \phi} g_\mathrm{S}^{\mu \rho} g_\mathrm{S}^{\nu \sigma} F_{\mu \nu} F_{\rho \sigma} - e^{- \phi} U(\phi, \Lambda) \right),
\end{equation}
    where the total derivative terms have been omitted. The corresponding Noether potential~\cite{LopesCardoso:1999cv, Parikh:2009qs} for the Wald entropy takes the following form:
\begin{equation}
    \mathcal{J}_\mathrm{S}^{\mu \nu} = -2 P_\mathrm{S}^{\mu\nu\rho\sigma} \nabla^\mathrm{S}_\rho \xi^\mathrm{S}_\sigma + 4 \xi^\mathrm{S}_\sigma \nabla^\mathrm{S}_\rho P_\mathrm{S}^{\mu\nu\rho\sigma},
\end{equation}
    where
\begin{equation}
    P_\mathrm{S}^{\mu\nu\rho\sigma} = \frac{\partial \mathcal{L}_\mathrm{S}}{\partial R^\mathrm{S}_{\mu\nu\rho\sigma}} = \frac{e^{-\phi}}{2} \left( g_\mathrm{S}^{\mu\rho} g_\mathrm{S}^{\nu\sigma} - g_\mathrm{S}^{\mu\sigma} g_\mathrm{S}^{\nu\rho} \right),
\end{equation}
    and the non-vanishing components of the Noether potential are found to be
\begin{equation}
    \mathcal{J}_\mathrm{S}^{tr} = -\mathcal{J}_\mathrm{S}^{rt} = e^{-2 \left( \phi + \frac{\phi_\infty}{4} \right)} \left( \frac{f'}{f} + 3 \phi' \right) g.
\end{equation}
    The binormal surface element is given by
\begin{equation}
    \mathrm{d}\sigma^\mathrm{S}_{\mu\nu} = \frac{1}{2} \left( n_\mu u_\nu - u_\mu n_\nu \right) \mathrm{d}\mathcal{A}_\mathrm{S},
\end{equation}
    where $u^\mu$ is a timelike unit normal vector and $n^\mu$ is a spacelike unit normal vector. The Wald entropy in the string frame is expressed as follows:
\begin{equation}
    S_\mathrm{S} = \frac{1}{8 \kappa_\mathrm{S}} \int_\Sigma \mathcal{J}_\mathrm{S}^{\rho\sigma} \mathrm{d}\sigma^\mathrm{S}_{\rho\sigma} = \pi \mathcal{R}(r_\mathrm{h})^2,
\end{equation}
    where $\Sigma$ denotes a spacelike cross-section of the Killing horizon. In contrast to the Hawking temperature, which exhibits frame dependence in terms of the asymptotic value of the dilaton field, the contribution of $\phi_\infty$ is canceled by the surface gravity in the denominator, and the expression is independent of the choice of frame. Thus, the Wald entropy in the string frame coincides with that in the Einstein frame as well as with the Bekenstein--Hawking entropy in the Einstein frame derived from the horizon area. Accordingly, for the dilatonic RN--AdS black hole, the Einstein and string frames have the same entropy, which is given by
\begin{equation}
    S = \pi r_\mathrm{h}^2 \left( 1 - \frac{r_-}{r_\mathrm{h}} \right)^\frac{2 \alpha^2}{1 + \alpha^2}.
\end{equation}
    Thus, both the Hawking temperature and black hole entropy are frame-independent in the Einstein and string frames. Consequently, the first law and the free energy are expressed by identical equations in both frames
\begin{equation}
    \mathrm{d}M = T_\mathrm{H} \mathrm{d}S + \Phi \mathrm{d}Q, \qquad F = M - T_\mathrm{H} S.
\end{equation}

\subsection{Lyapunov Exponents}
    To examine the dynamical properties of particle motion in the dilatonic RN--AdS black hole, we study the effective potentials and the Lyapunov exponents, the latter of which quantify orbital instability. The particle dynamics are described by a Polyakov-type Lagrangian for an electrically charged particle~\cite{Kan:2021blg, Gwak:2022xje, Park:2023lfc, Lee:2025vih, Lee:2025ias}, given by:
\begin{equation}
    \mathcal{L} = \frac{1}{2 \epsilon(s)} \left( \frac{\mathrm{d}X}{\mathrm{d}s} \right)^2 - \frac{\epsilon(s)}{2} m^2 + q A_\mu \frac{\mathrm{d}X^\mu}{\mathrm{d}s},
\end{equation}
    where $m$ and $q$ represent the mass and electric charge of the particle, respectively; $\epsilon$ is an auxiliary field; $s$ parametrizes the particle trajectory; and $X^\mu = ( t(s), \, r(s), \, \theta(s), \, \varphi(s) )$ denotes the spacetime coordinates. The static gauge $t = s$ is imposed, and the particle motion is restricted to the equatorial plane $(\theta = \pi/2)$. With the metric~\eqref{eq:metric_Einstein}, the resulting Lagrangians in the Einstein and string frames are as follows:
\begin{equation}
    \mathcal{L}_\mathrm{E} = \left. \mathcal{L}_\mathrm{S} \right|_{\phi = 0}, \qquad \mathcal{L}_\mathrm{S} = \frac{e^{\phi}}{2 \epsilon} \left( \frac{\dot{r}^2}{g} + \mathcal{R}^2 \dot{\varphi}^2  - f \right) - \frac{\epsilon}{2} m^2 - \frac{q Q}{r},
\end{equation}
    where $|_{\phi = 0}$ represents the Einstein frame limit derived by imposing the condition $\phi = 0$ on the string frame expression. Variation with respect to the auxiliary field produces the constraint $\dot{X}^2 = - \epsilon^2 m^2$ in each frame. Owing to the absence of explicit $\varphi$ dependence in the Lagrangian, the conserved angular momentum $L \equiv \partial \mathcal{L} / \partial \dot{\varphi}$ can be defined, with its explicit forms in the Einstein and string frames given by
\begin{equation}
    L_\mathrm{E} = \left. L_\mathrm{S} \right|_{\phi = 0}, \qquad L_\mathrm{S} = \frac{e^\phi \mathcal{R}^2}{\epsilon_\mathrm{S}} \dot{\varphi}.
\end{equation}
    Notably, the conserved angular momentum explicitly depends on the choice of frame under the conformal transformation. However, because the auxiliary fields are defined independently in each frame, they can be chosen consistently to yield the same conserved angular momentum in both frames, {\it i.e.}, $L_\mathrm{E} = L_\mathrm{S} \equiv L$. Adopting the same conserved quantity in both frames facilitates a direct comparison of the corresponding particle dynamics between the Einstein and string frames. The auxiliary fields are then expressed as follows
\begin{equation}
    \epsilon_\mathrm{E}^2 = \left. \epsilon_\mathrm{S}^2 \right|_{\phi = 0}, \qquad \epsilon_\mathrm{S}^2 = - \frac{e^{2 \phi} \left( \frac{\dot{r}^2}{g} - f \right)}{e^\phi m^2 + \frac{L^2}{\mathcal{R}^2}}.
\end{equation}
    Inserting the expressions for the conserved angular momentum and auxiliary field yields the effective Lagrangian $\mathcal{L}^\mathrm{eff} \equiv \mathcal{L} - L \dot{\varphi}$, which is expressed explicitly in the Einstein and string frames as follows:
\begin{align}
    \mathcal{L}^\mathrm{eff}_\mathrm{E} = \left. \mathcal{L}^\mathrm{eff}_\mathrm{S} \right|_{\phi = 0}, \qquad \mathcal{L}^\mathrm{eff}_\mathrm{S} = - \sqrt{e^\phi m^2 + \frac{L^2}{\mathcal{R}^2}} \sqrt{- \frac{\dot{r}^2}{g} + f} - \frac{q Q}{r}.
\end{align}
    In the non-relativistic limit $(\dot{r} \ll 1)$, this reduces to
\begin{equation}
    \mathcal{L}^\mathrm{eff}_\mathrm{E} = \frac{1}{2} K_\mathrm{E}(r) \dot{r}^2 - V_\mathrm{E}(r) + \mathcal{O}(\dot{r}^4), \qquad \mathcal{L}^\mathrm{eff}_\mathrm{S} = \frac{1}{2} K_\mathrm{S}(r) \dot{r}^2 - V_\mathrm{S}(r) + \mathcal{O}(\dot{r}^4),
\end{equation}
    where
\begin{align}
    K_\mathrm{E}(r) &= \frac{1}{f^{1/2} g} \sqrt{m^2 + \frac{L^2}{\mathcal{R}^2}}, \qquad\;\:\:\, V_\mathrm{E}(r) = f^{1/2} \sqrt{m^2 + \frac{L^2}{\mathcal{R}^2}} + \frac{q Q}{r},
    \label{eq:kinetic_potential_Einstein}
    \\
    K_\mathrm{S}(r) &= \frac{1}{f^{1/2} g} \sqrt{e^\phi m^2 + \frac{L^2}{\mathcal{R}^2}}, \qquad V_\mathrm{S}(r) = f^{1/2} \sqrt{e^\phi m^2 + \frac{L^2}{\mathcal{R}^2}} + \frac{q Q}{r}.
    \label{eq:kinetic_potential_string}
\end{align}
    To analyze the behavior near the local maximum of the effective potential, the particle is considered initially at rest at $r_0$, where $V'(r_0) = 0$ and $V''(r_0) < 0$. Introducing a small deviation $\delta r$ in the vicinity of $r_0$, the effective Lagrangian can be expanded as follows:
\begin{equation}
    \mathcal{L}^\mathrm{eff} = \frac{1}{2} K(r_0) \left( \delta \dot{r}^2 + \lambda^2 \delta r^2 \right),
\end{equation}
    where $\lambda$ denotes the Lyapunov exponent, and constant terms have been omitted. The Lyapunov exponents in the Einstein and string frames are then obtained as follows:
\begin{equation}
    \lambda_\mathrm{E} \equiv \sqrt{- \frac{V_\mathrm{E}''(r_0)}{K_\mathrm{E}(r_0)}}, \qquad \lambda_\mathrm{S} \equiv \sqrt{- \frac{V_\mathrm{S}''(r_0)}{K_\mathrm{S}(r_0)}}.
\end{equation}
    For the dilatonic RN--AdS black hole with the condition $V'(r_0) = 0$, the Lyapunov exponents reduce to
\begin{equation}
    \lambda_\mathrm{E} = \left. \lambda_\mathrm{S} \right|_{\phi = 0}, \qquad \lambda_\mathrm{S} = \left. f \sqrt{\frac{K_\mathrm{S}''}{K_\mathrm{S}} + \frac{2 f''}{f} - \frac{6 (f')^2}{f^2} + \frac{2 q Q \left( 2 r f' + f \right)}{r^3 f^3 K_\mathrm{S}}} \right|_{r = r_0}.
\end{equation}
    The resulting expressions facilitate a direct comparison between the Einstein and string frames in terms of dynamical stability. In contrast to the thermodynamic quantities, the Lyapunov exponent exhibits explicit frame dependence induced by dilaton coupling, thereby capturing frame-dependent features. The distinction vanishes in the massless limit $(m = 0)$, while for a massive particle, the dilaton coupling effectively modifies the particle mass, giving rise to frame dependence~\cite{Lee:2025ias}.
    
    Thus, thermodynamic quantities, including the Hawking temperature and black hole entropy, remain constant under conformal transformations between the Einstein and string frames, whereas the Lyapunov exponent exhibits non-trivial frame dependence. This contrast highlights a distinction between the equilibrium and dynamical aspects of the system in the presence of non-minimal dilaton coupling. The analytical results are corroborated by numerical analysis in Sec.~\ref{sec:numerical_results}.

\section{Numerical Results} \label{sec:numerical_results}
    This section presents the numerical results for the dilatonic RN--AdS black hole, focusing on its thermodynamic and dynamical properties. The free energy is used to study phase transitions from a thermodynamic perspective, while the Lyapunov exponent is used to study the same transitions from a dynamical perspective. Furthermore, the frame dependence and the effects of the dilaton coupling parameter $\alpha$ are examined. Using the equation of motion for the gauge field, the black hole charge in Eq.~\eqref{eq:black_hole_charge} is directly obtained from the horizon parameters $r_+$ and $r_-$. Thus, a dimensionless formulation is introduced by rescaling all quantities by $Q$
\begin{equation}
    \tilde{\Lambda} \equiv Q^2 \Lambda, \quad \tilde{M} \equiv \frac{M}{Q}, \quad \tilde{F} \equiv \frac{F}{Q}, \quad \tilde{T} \equiv Q T, \quad \tilde{\lambda} \equiv Q \lambda, \quad \tilde{r} \equiv \frac{r}{Q}, \quad \tilde{m} \equiv \frac{m}{Q}, \quad \tilde{q} \equiv \frac{q}{Q}, \quad \tilde{L} \equiv \frac{L}{Q^2}.
\end{equation}
    In the present analysis, the black hole charge is fixed to $Q = 1$, which is equivalent to expressing all quantities in terms of dimensionless variables rescaled by $Q$ without loss of generality, and the tilde symbols are omitted for simplicity. The mass of the probe particle is set to $m = 1$ for numerical convenience.

\subsection{Asymptotically Flat Spacetime}
    First, the asymptotically flat case is considered to investigate the phase structure of the dilatonic RN black holes.
\begin{figure}[H]
    \centering
    \subfigure[$\alpha = 0.1$]{\includegraphics[width=5.5cm]{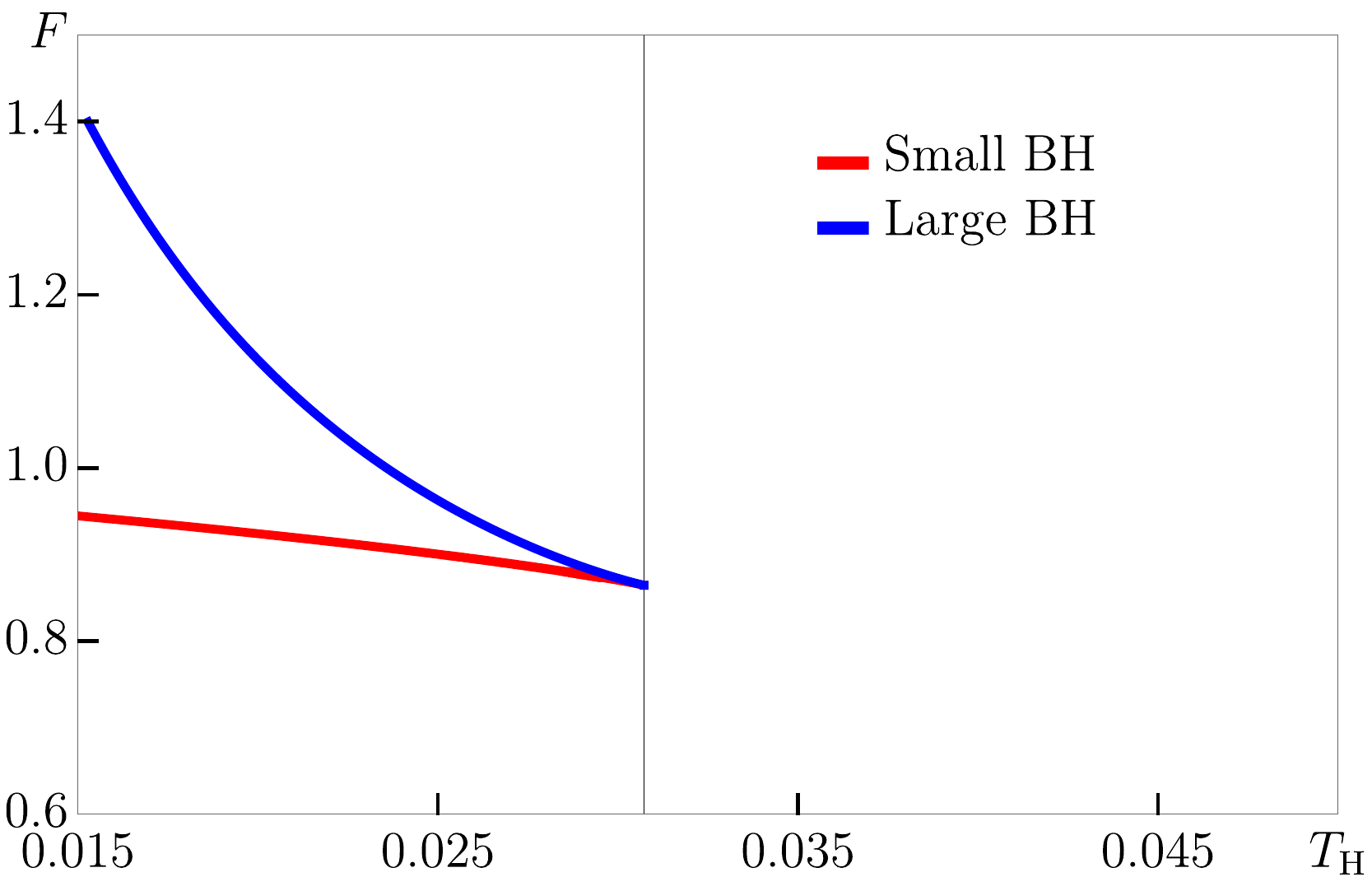}}
    \subfigure[$\alpha = 0.5$]{\includegraphics[width=5.5cm]{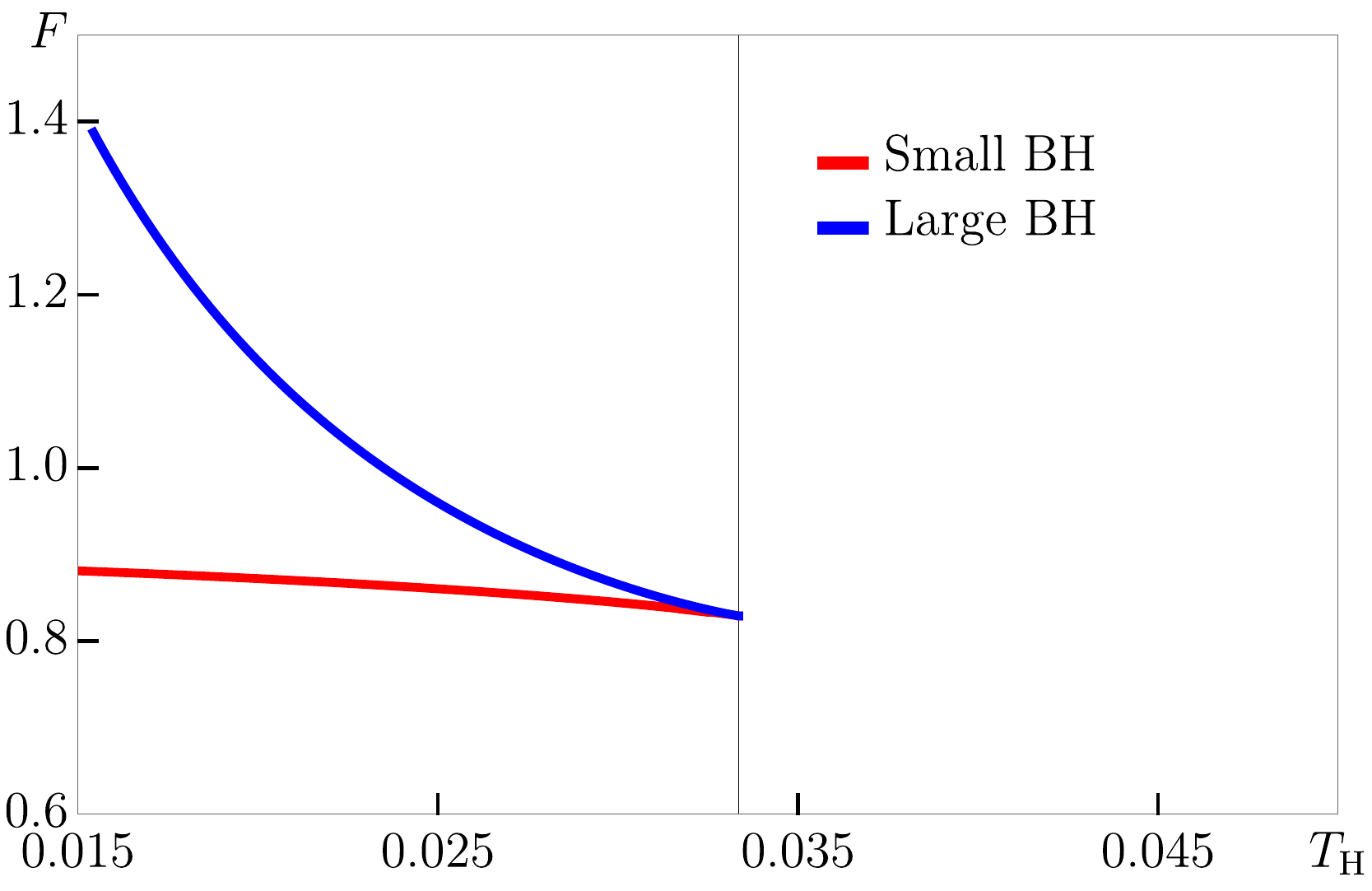}}
    \subfigure[$\alpha = 0.9$]{\includegraphics[width=5.5cm]{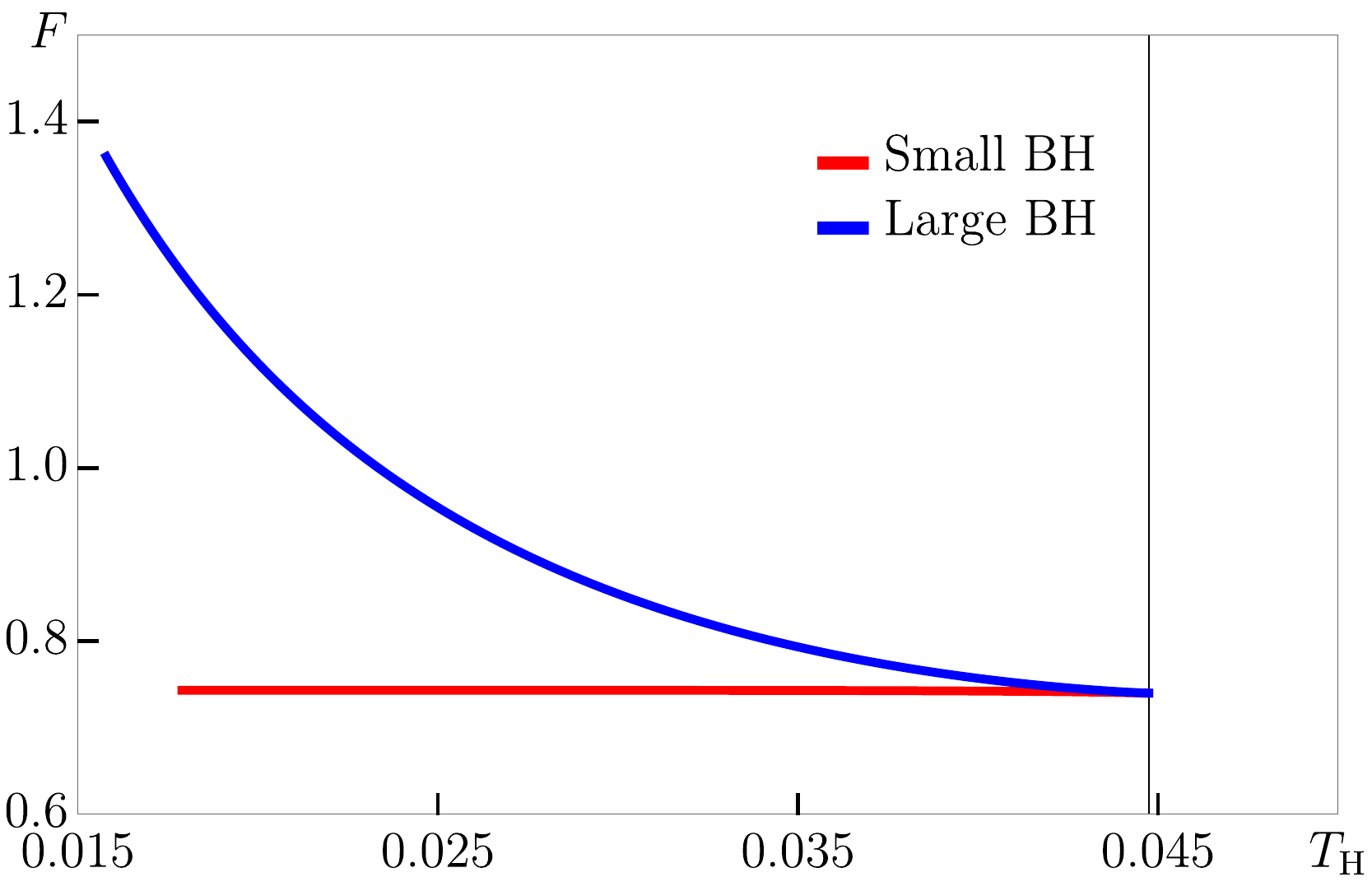}}
    \caption{Free energy as a function of the Hawking temperature in the dilatonic RN black hole at different values of the dilaton coupling $\alpha$.}
    \label{fig:phase_structure_flat}
\end{figure}
    Fig.~\ref{fig:phase_structure_flat} presents the free energy as a function of the Hawking temperature in the asymptotically flat case for $\alpha = 0.1$, $0.5$, and $0.9$, in which the small and large black hole branches are represented by red and blue lines, respectively. The cusp structure indicates a multi-valued behavior of the free energy as a function of temperature. The cusp appears at the maximum temperature $T_\mathrm{max}$, given by
\begin{equation}
    T_\mathrm{max} = \frac{2^\frac{2}{1 + \alpha^2}}{8 \pi \sqrt{3 - \alpha^2} |Q|} \left( \frac{1 - \alpha^2}{3 - \alpha^2} \right)^\frac{1 - \alpha^2}{1 + \alpha^2},
\end{equation}
    with $T_\mathrm{max} = 0.0307$,~$0.0334$, and~$0.0447$ for $\alpha = 0.1$,~$0.5$, and~$0.9$, respectively. As the dilaton coupling parameter $\alpha$ increases, the maximum temperature also increases. The specific heat $C_Q$ at fixed charge is defined as follows:
\begin{equation}
    C_Q \equiv T_\mathrm{H} \left( \frac{\partial S}{\partial T_\mathrm{H}} \right)_Q = - 2 \pi r_+^2 \left[ \frac{r_+ - \left( \frac{1 - \alpha^2}{1 + \alpha^2} \right) r_-}{r_+ - \left( \frac{3 - \alpha^2}{1 + \alpha^2} \right) r_-} \right] \left( 1 - \frac{r_-}{r_+} \right)^\frac{2 \alpha^2}{1 + \alpha^2}.
\end{equation}   
    The specific heat is positive for $r_+ < (3 - \alpha^2) / (1 + \alpha^2) r_-$ and negative for $r_+ > (3 - \alpha^2) / (1 + \alpha^2) r_-$. It diverges at $r_+ = (3 - \alpha^2) / (1 + \alpha^2) r_-$, which coincides with the maximum temperature. The positive specific heat corresponds to the small black hole branch (red line), whereas the negative specific heat corresponds to the large black hole branch (blue line). While the free energy profile exhibits the coexistence of multiple black hole branches, the large black hole branch is thermodynamically unstable, as suggested by the specific heat.
    
\begin{figure}[H]
    \centering
    \subfigure[$q = 2$, $L = 0.01$, $\alpha = 0.1$]{\includegraphics[width=5.5cm]{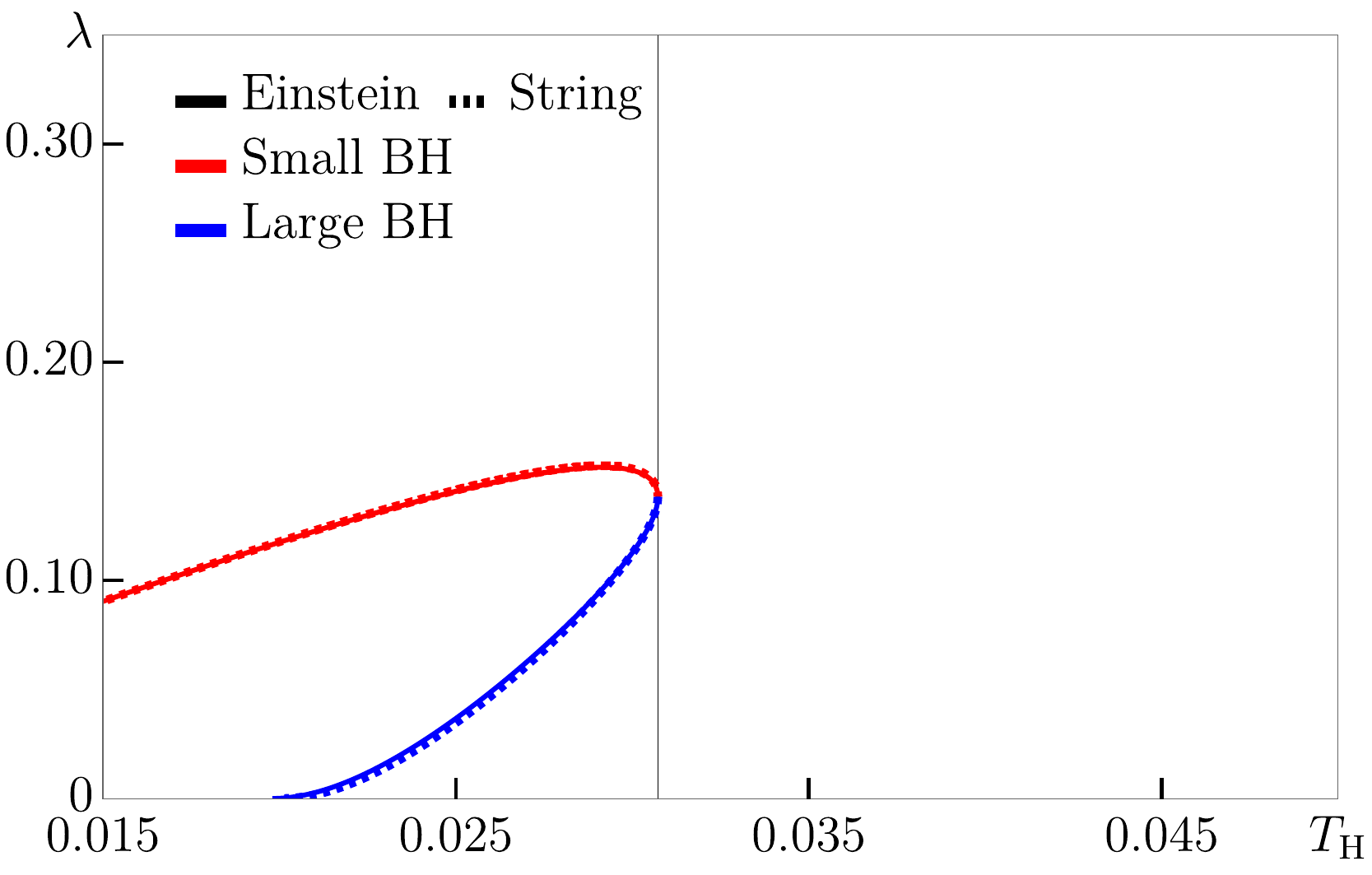}}
    \subfigure[$q = 2$, $L = 0.01$, $\alpha = 0.5$]{\includegraphics[width=5.5cm]{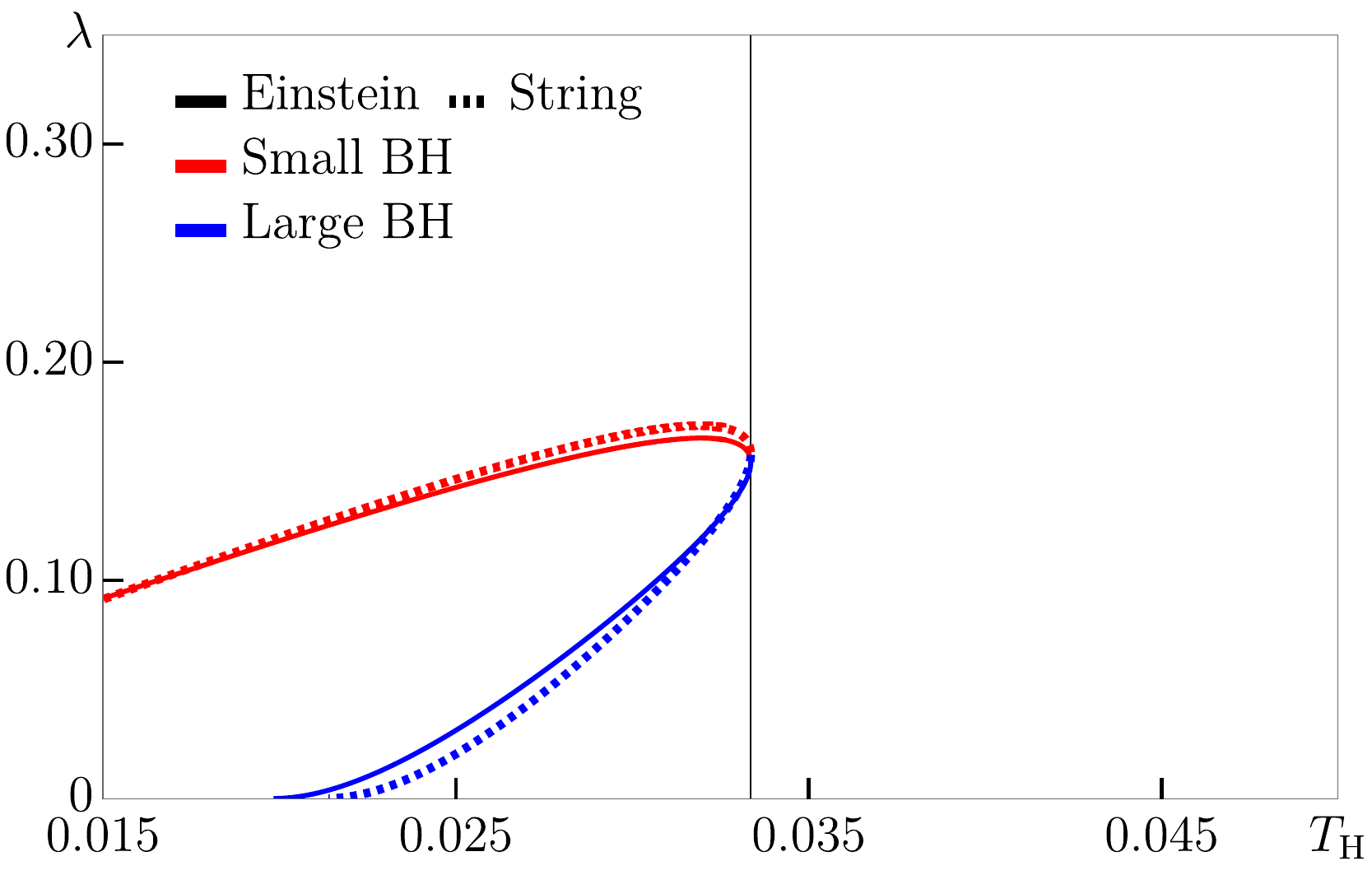}}
    \subfigure[$q = 2$, $L = 0.01$, $\alpha = 0.9$]{\includegraphics[width=5.5cm]{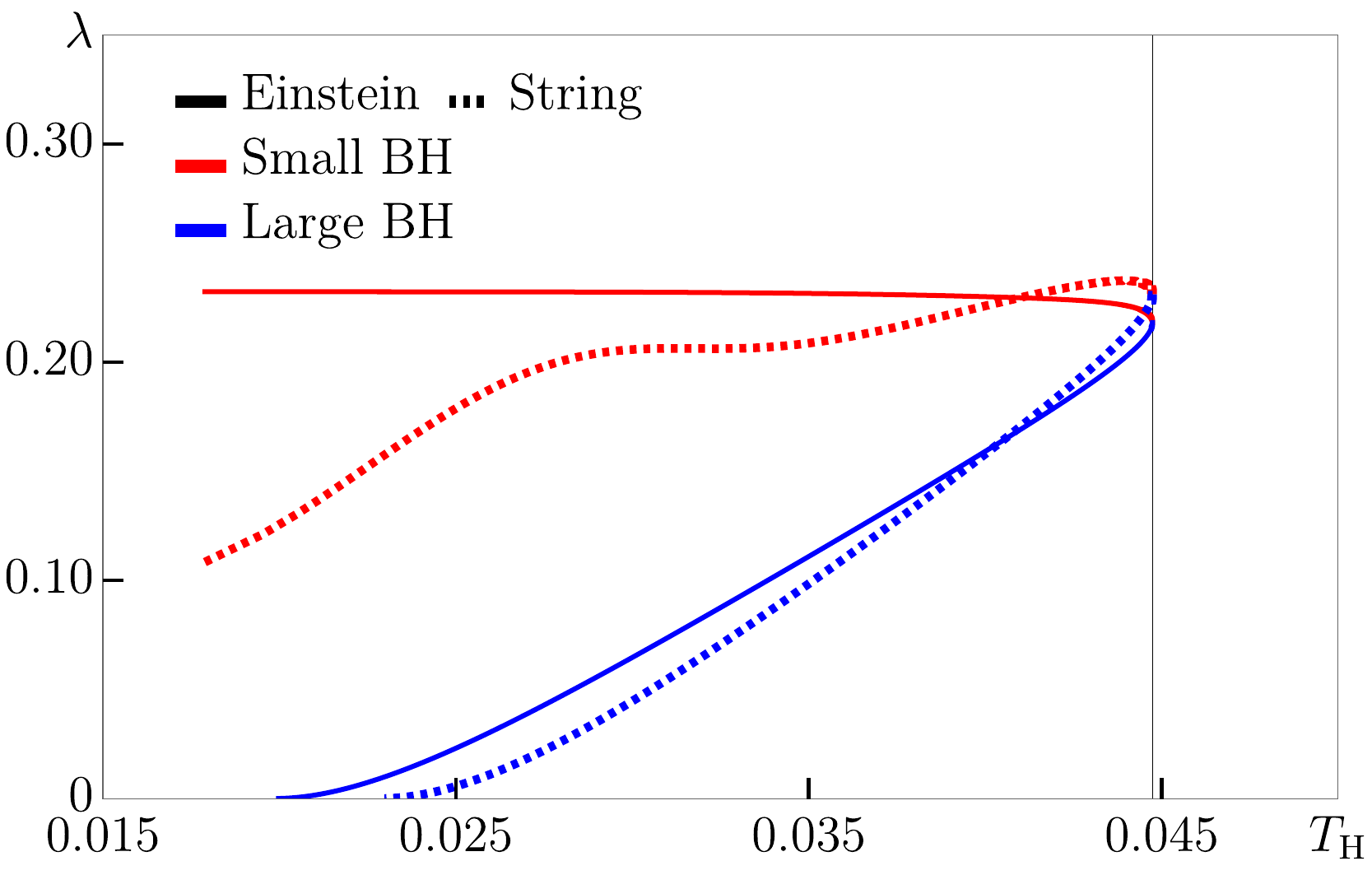}}
    \subfigure[$q = 20$, $L = 0.01$, $\alpha = 0.1$]{\includegraphics[width=5.5cm]{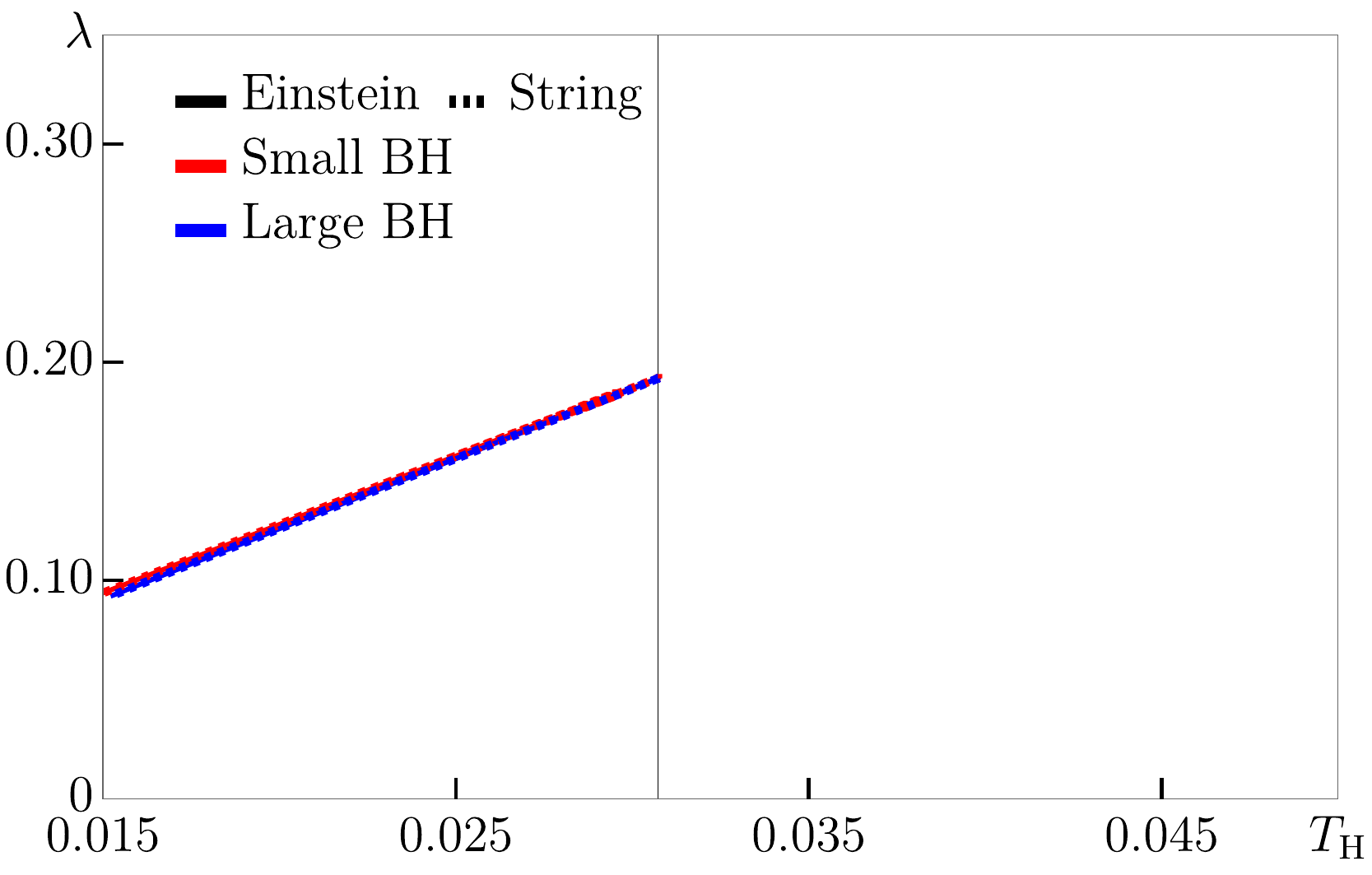}}
    \subfigure[$q = 20$, $L = 0.01$, $\alpha = 0.5$]{\includegraphics[width=5.5cm]{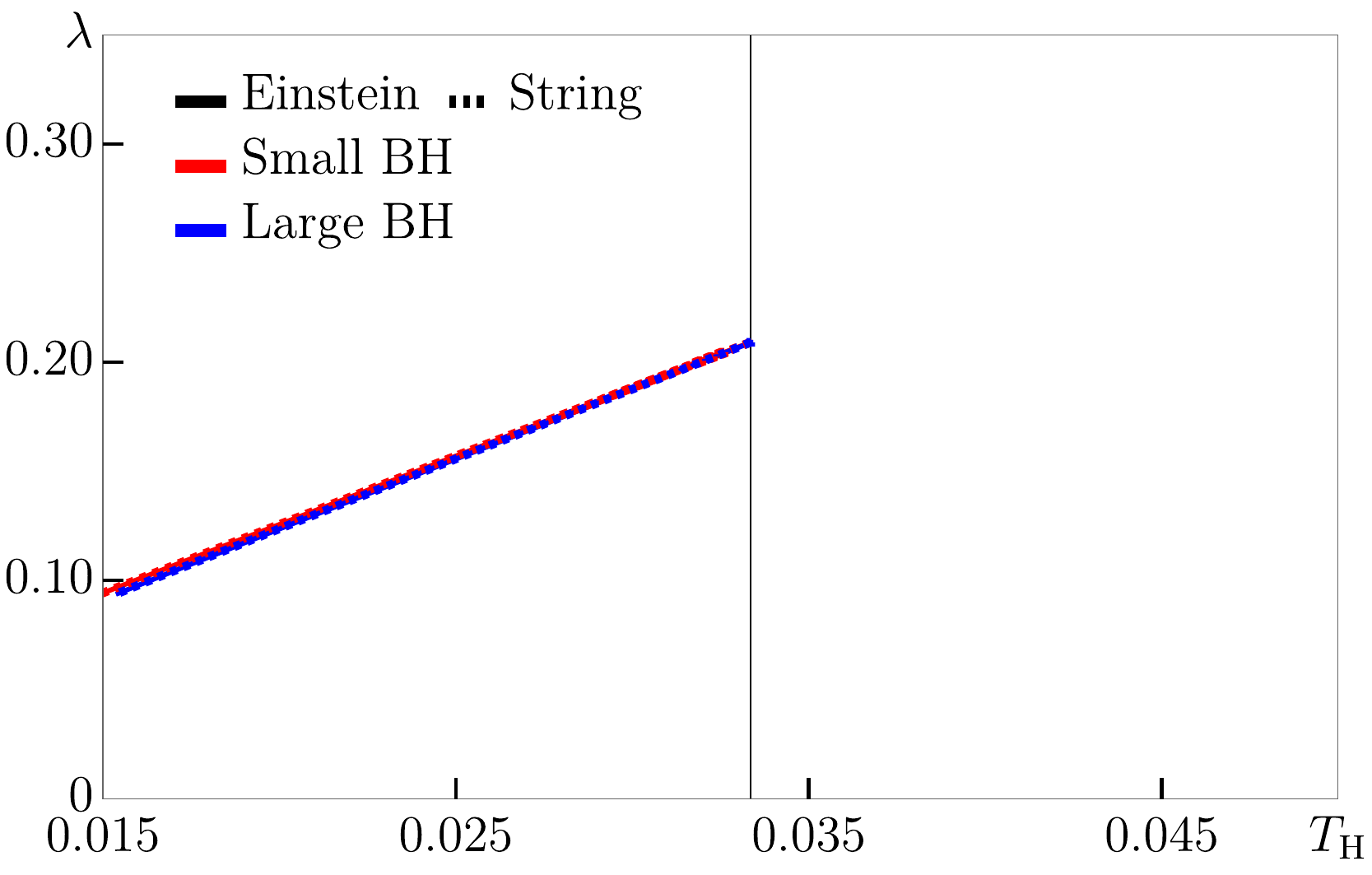}}
    \subfigure[$q = 20$, $L = 0.01$, $\alpha = 0.9$]{\includegraphics[width=5.5cm]{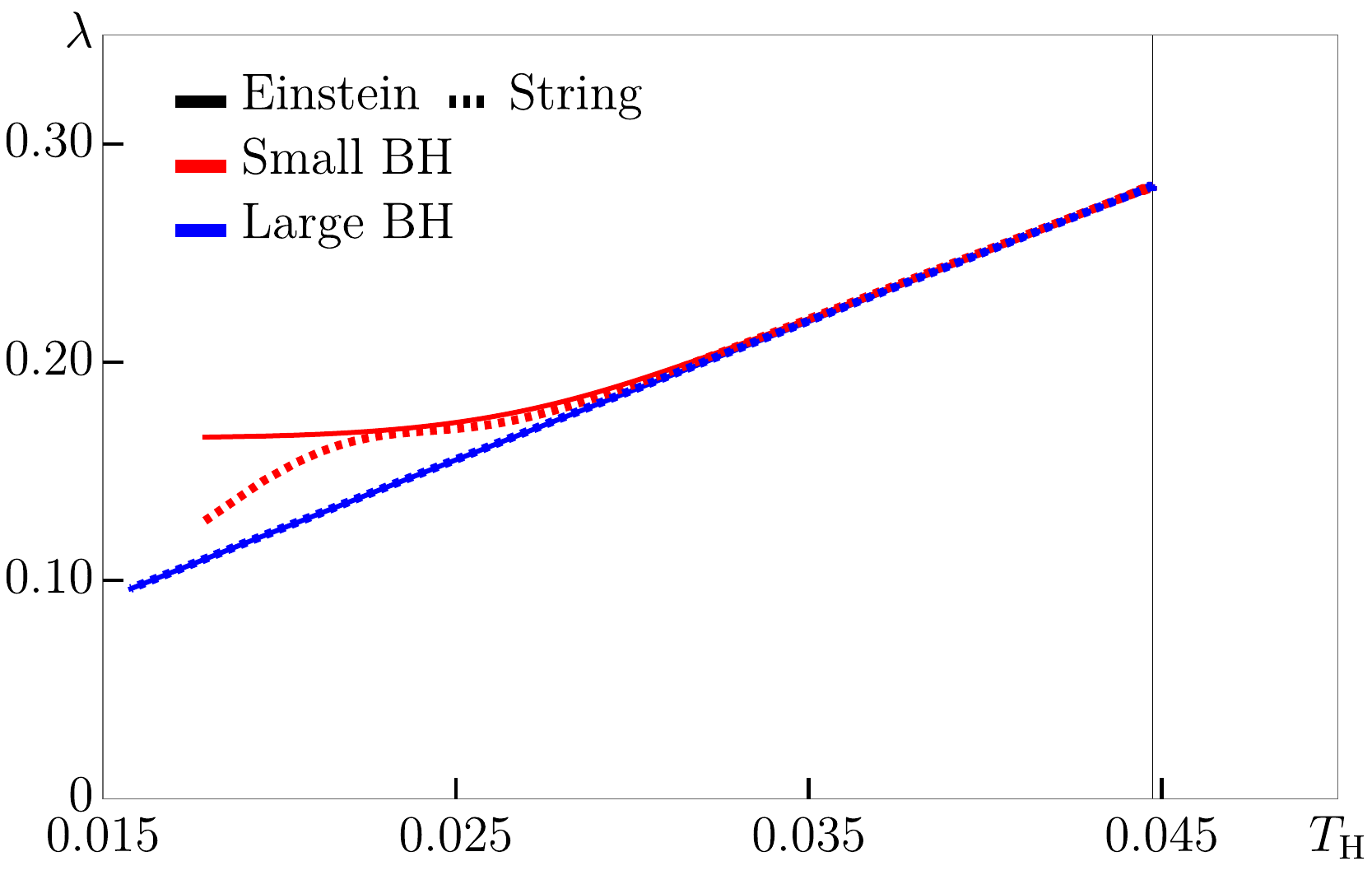}}
    \subfigure[$q = 2$, $L = 10$, $\alpha = 0.1$]{\includegraphics[width=5.5cm]{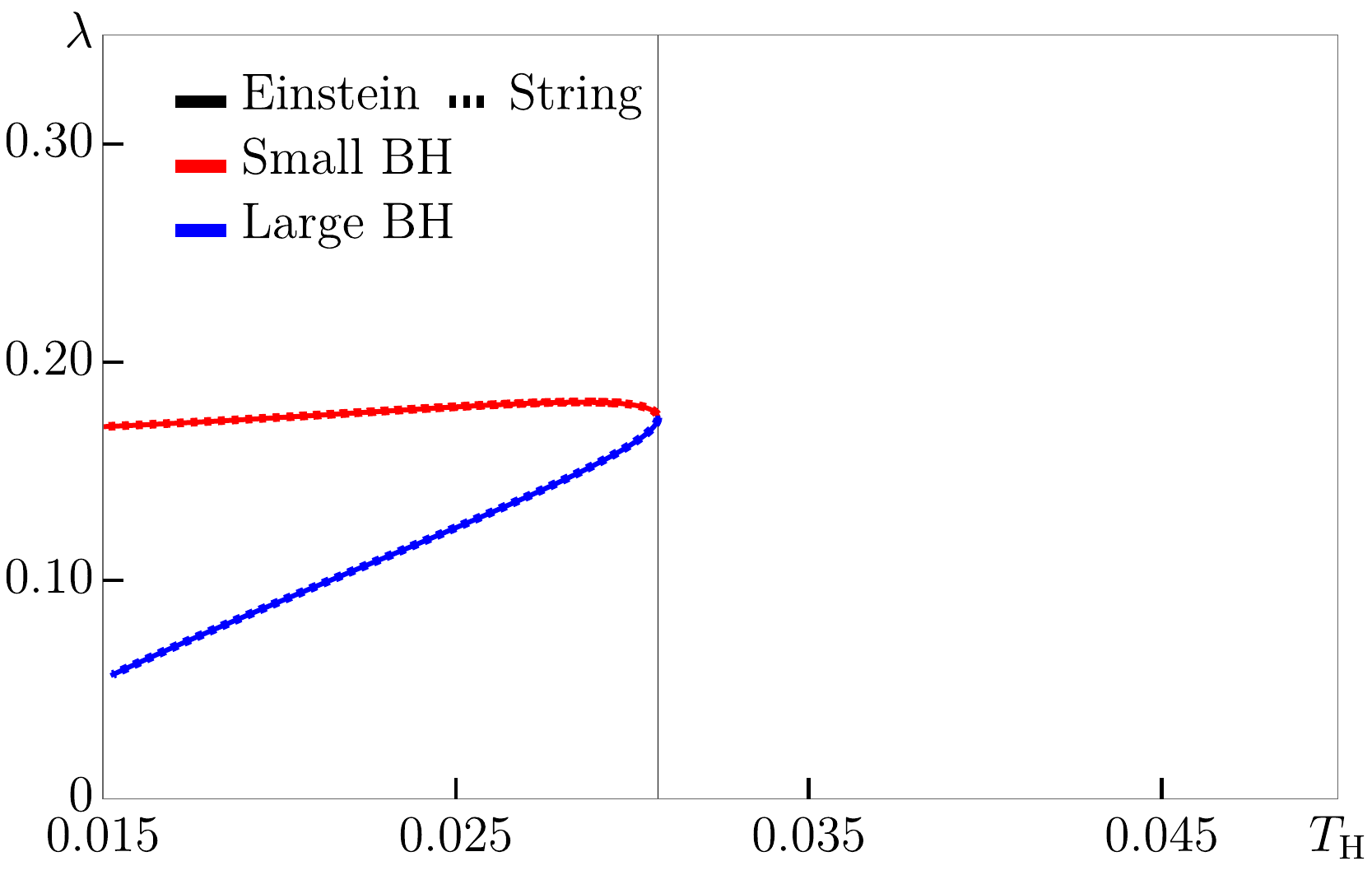}}
    \subfigure[$q = 2$, $L = 10$, $\alpha = 0.5$]{\includegraphics[width=5.5cm]{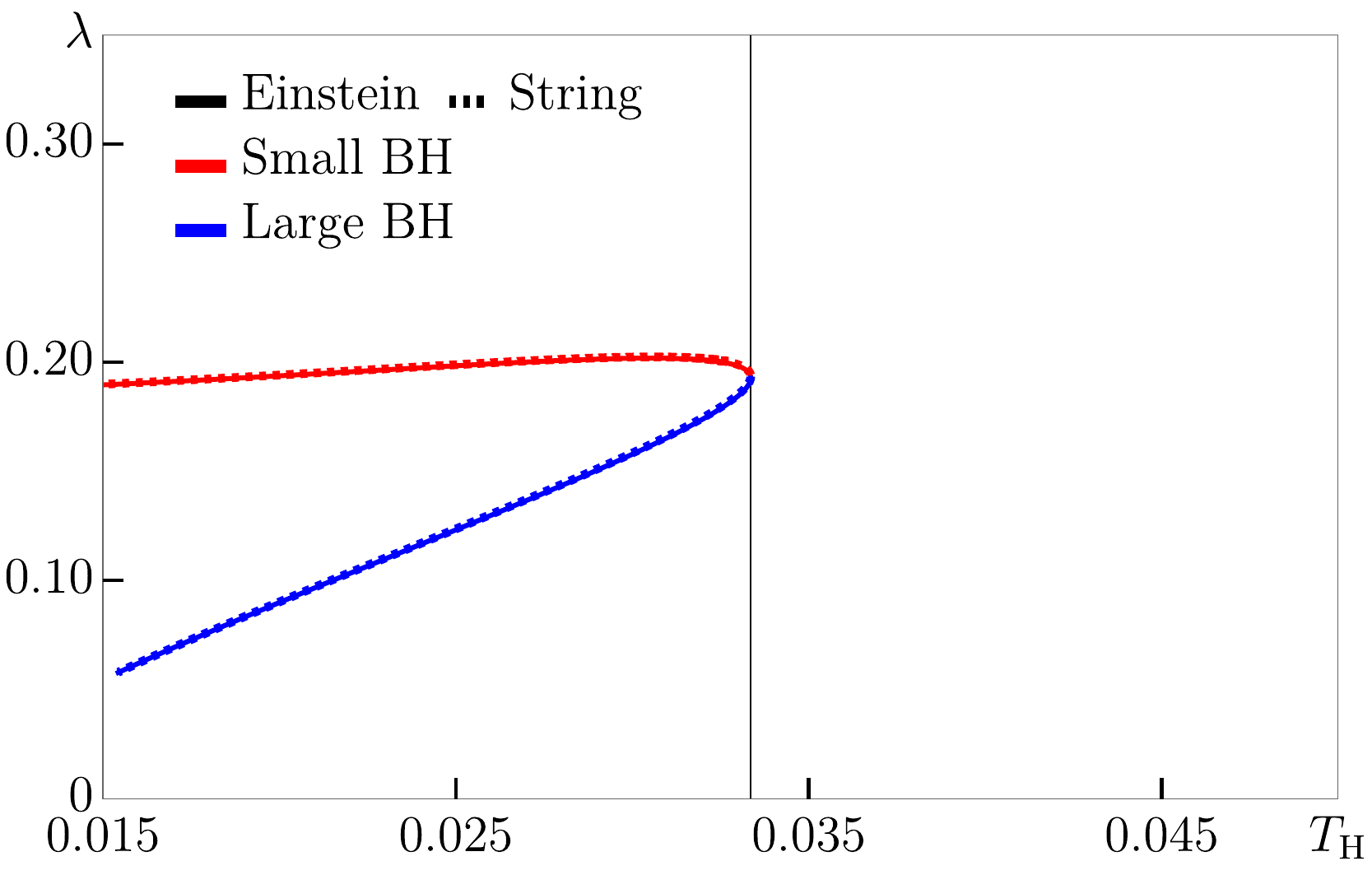}}
    \subfigure[$q = 2$, $L = 10$, $\alpha = 0.9$]{\includegraphics[width=5.5cm]{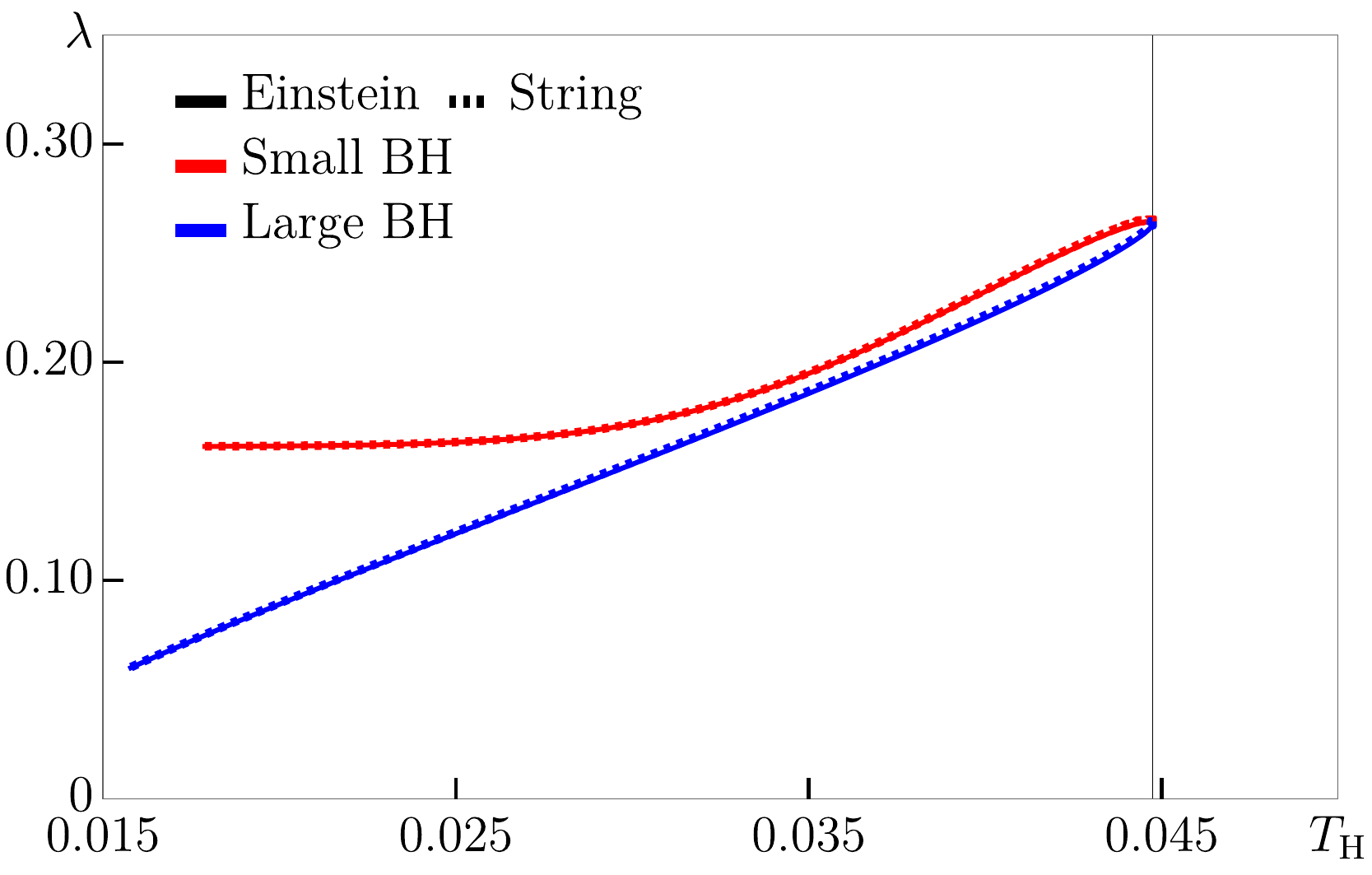}}
    \caption{Lyapunov exponents as a function of the Hawking temperature for the dilatonic RN black hole with $\alpha = 0.1$, $0.5$, $0.9$.}
    \label{fig:Lyapunov_exponent_flat}
\end{figure}
    Fig.~\ref{fig:Lyapunov_exponent_flat} illustrates the Lyapunov exponent as a function of the Hawking temperature for $\alpha = 0.1$, $0.5$, and $0.9$, considering various combinations of the particle charge $q$ and angular momentum $L$. Frame dependence is pronounced at $\alpha = 0.9$, while it is substantially reduced for smaller values of $\alpha$. Notably, the difference between the two frames is maximized for small values of $q$ and $L$. Eqs.~\eqref{eq:kinetic_potential_Einstein} and~\eqref{eq:kinetic_potential_string} suggest that the effective kinetic term and potential contain competing contributions from the dilaton field, angular momentum, and particle charge. For sufficiently small values of $q$ and $L$, the dilaton contribution dominates, giving rise to stronger frame dependence. Furthermore, the Lyapunov exponent exhibits a smooth cusp-like structure at the maximum Hawking temperature, paralleling the cusp observed in Fig.~\ref{fig:phase_structure_flat}. The numerical results demonstrate that while the location of the cusp-like structure is independent of the choice of frame, the magnitude of the Lyapunov exponent is highly frame-dependent for massive particles. For larger values of $q$ or $L$, the dilaton contribution becomes less significant, reducing the difference between the Einstein and string frames.

\subsection{Asymptotically AdS Spacetime}
    This subsection presents the numerical results for the dilatonic RN--AdS black hole in asymptotically AdS spacetime.
    
\begin{figure}[H]
    \centering
    \subfigure[$\Lambda = -0.01$, $\alpha = 0.1$]{\includegraphics[width=5.5cm]{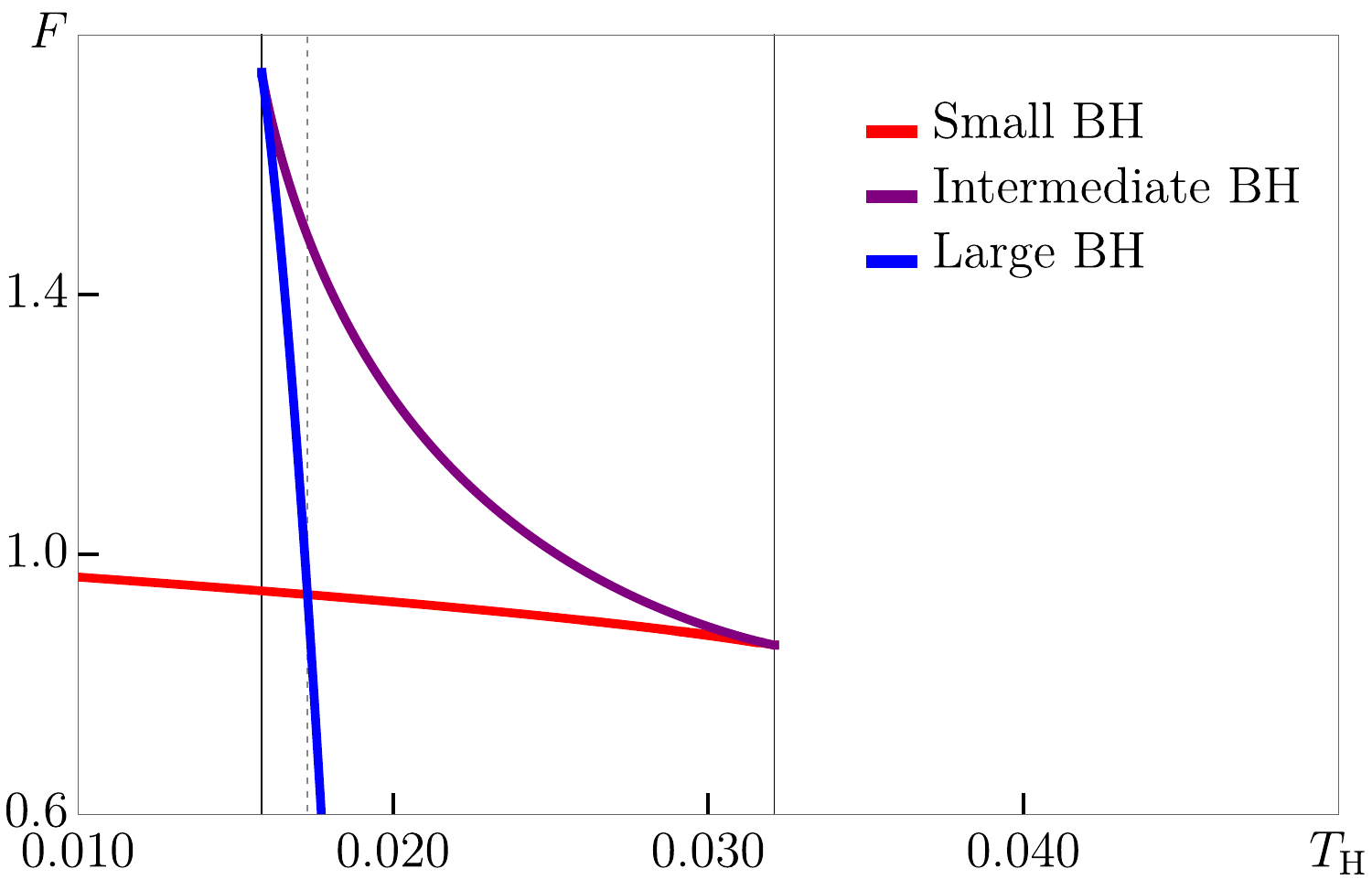}}
    \subfigure[$\Lambda = -0.01$, $\alpha = 0.5$]{\includegraphics[width=5.5cm]{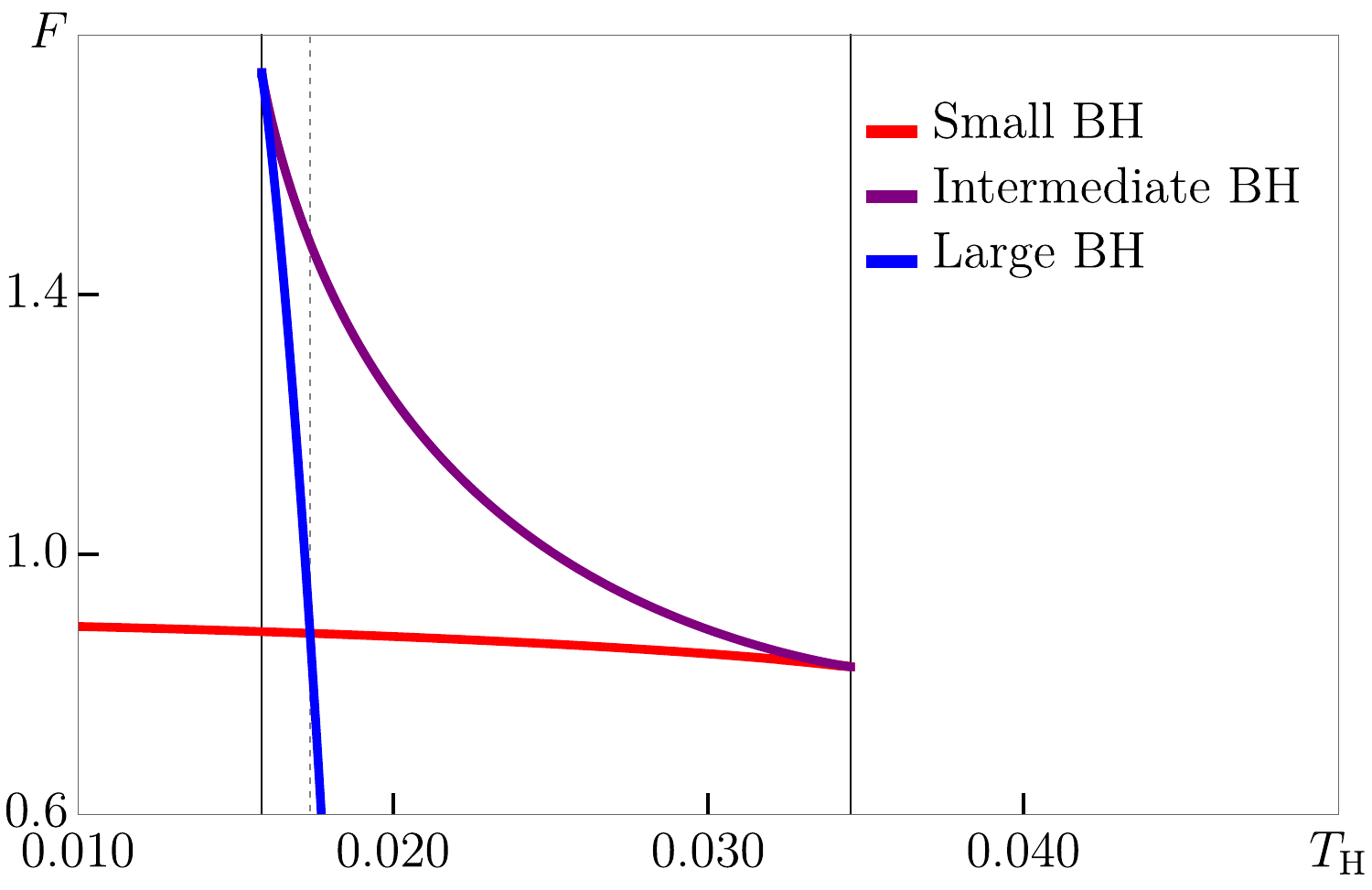}}
    \subfigure[$\Lambda = -0.01$, $\alpha = 0.9$ \label{subfig:c}]{\includegraphics[width=5.5cm]{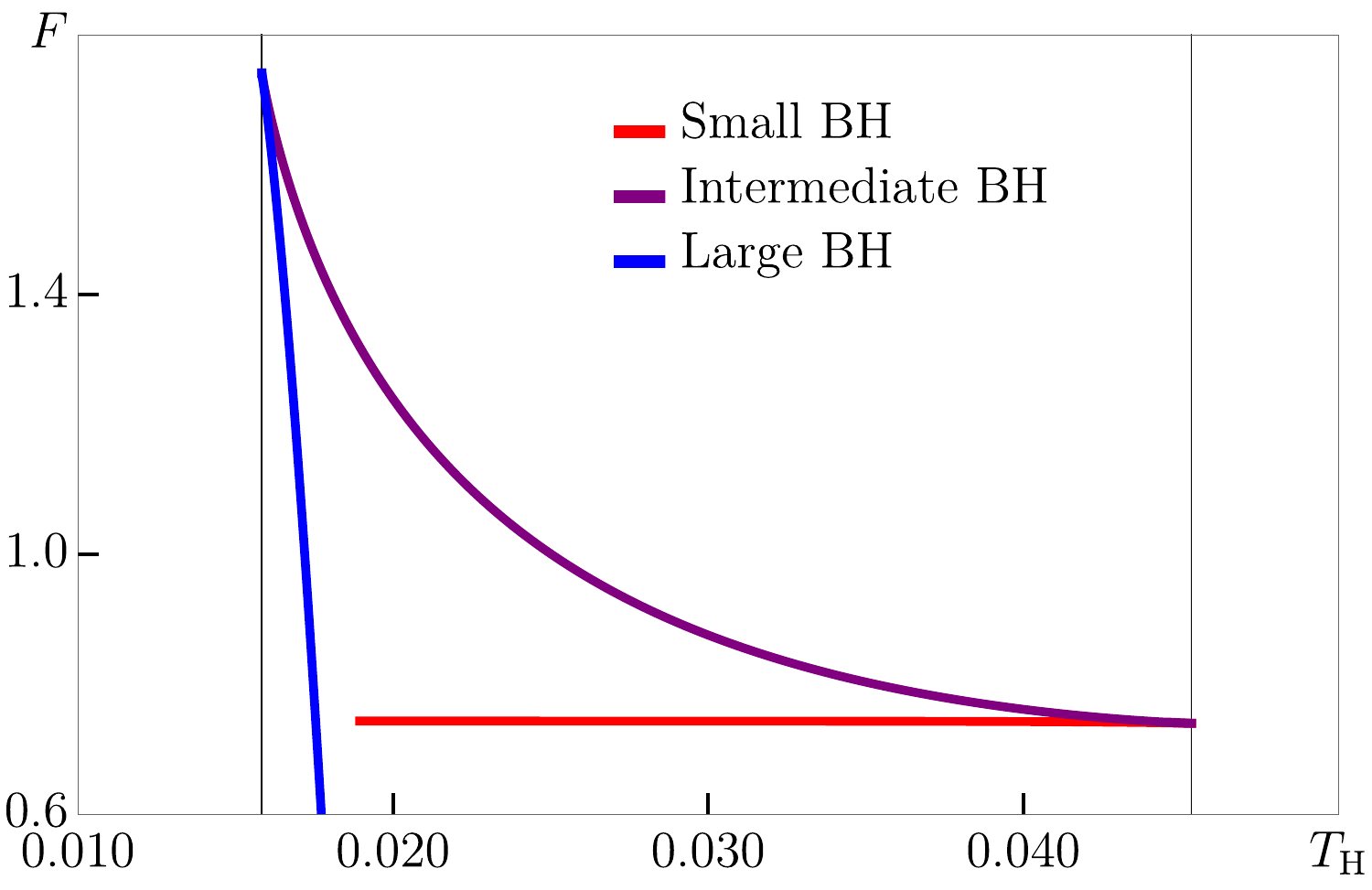}}
    \subfigure[$\Lambda = -0.05$, $\alpha = 0.1$]{\includegraphics[width=5.5cm]{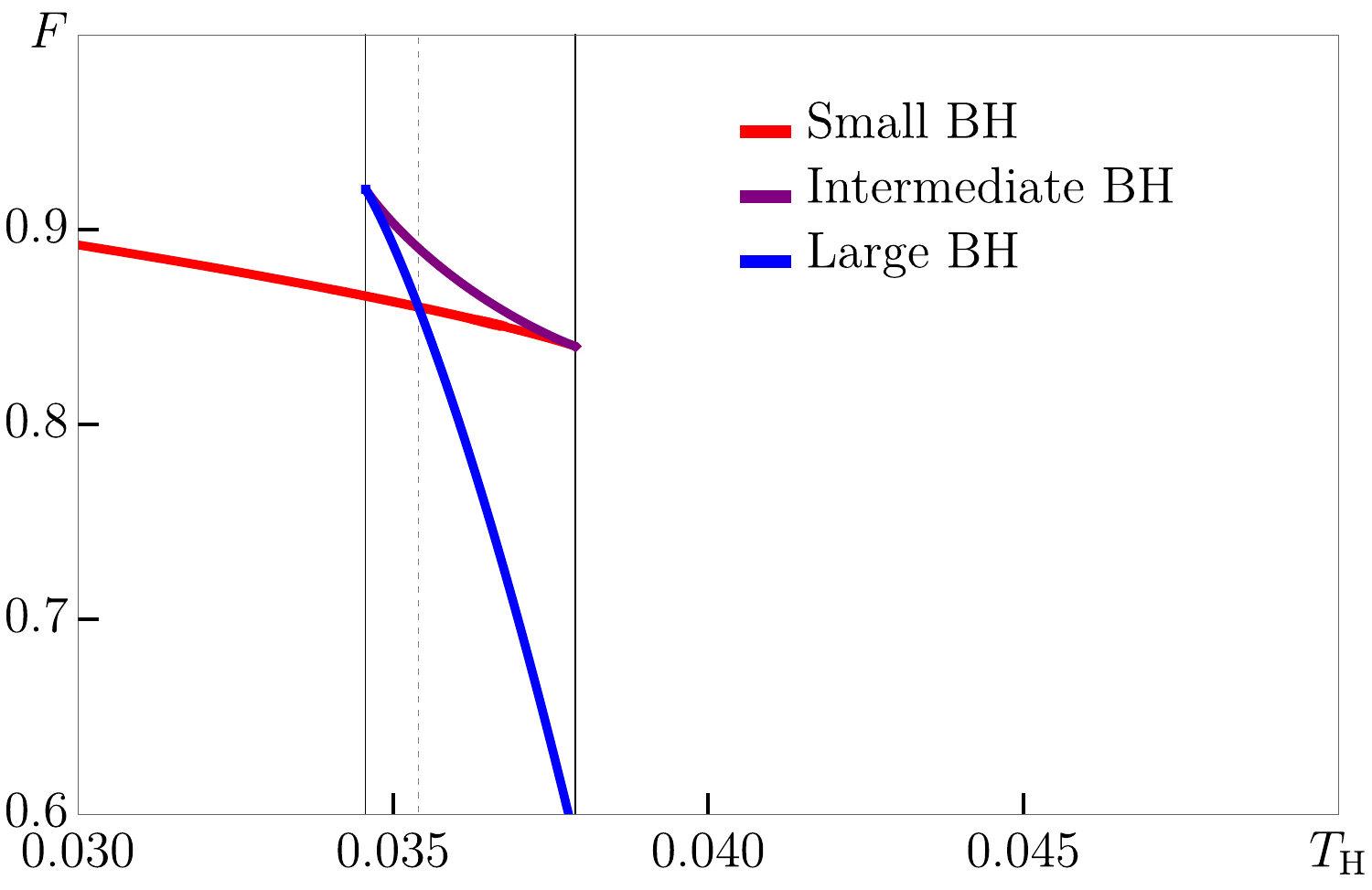}}
    \subfigure[$\Lambda = -0.05$, $\alpha = 0.5$]{\includegraphics[width=5.5cm]{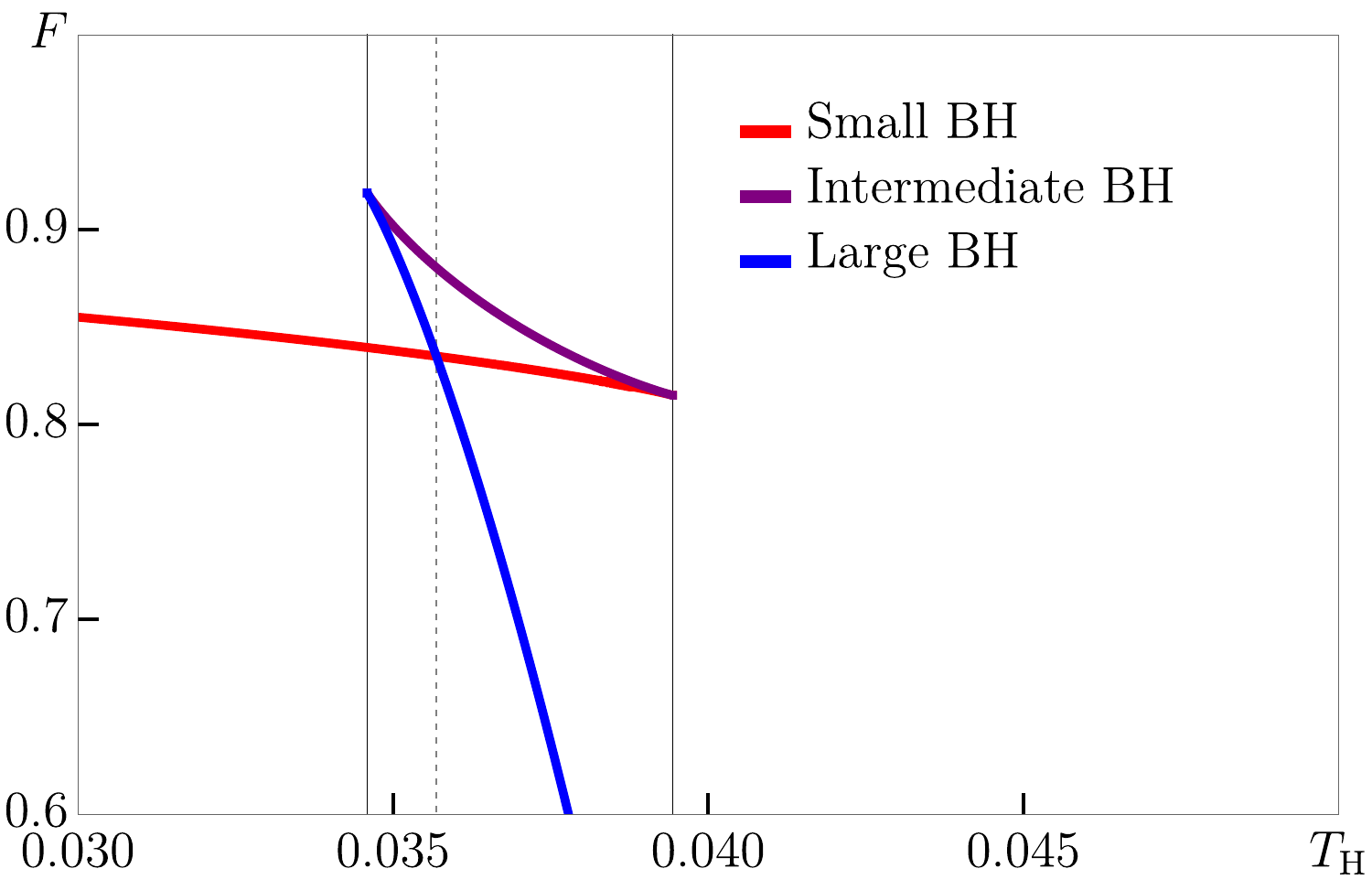}}
    \subfigure[$\Lambda = -0.05$, $\alpha = 0.9$]{\includegraphics[width=5.5cm]{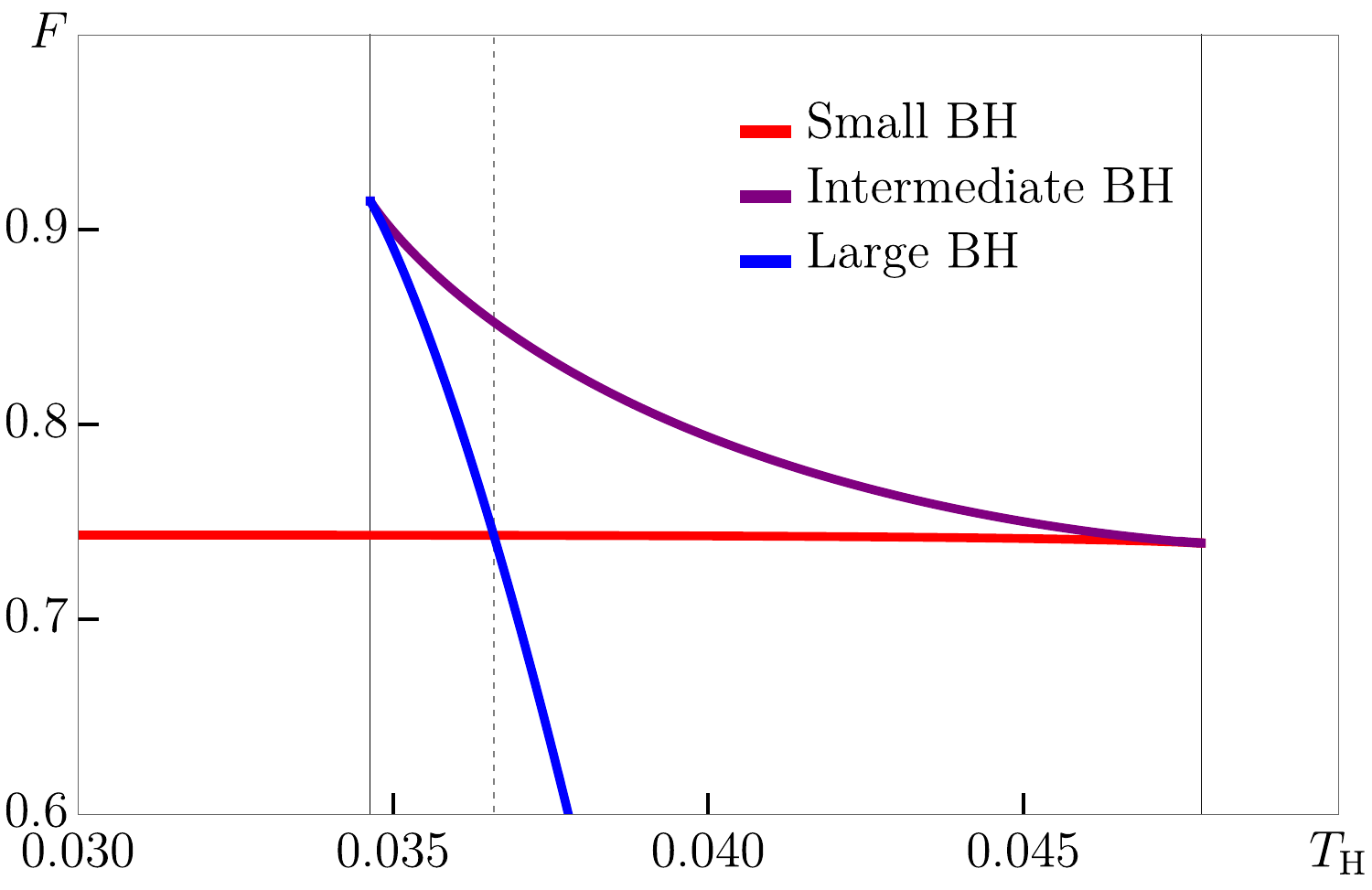}}
    \subfigure[$\Lambda = -0.09$, $\alpha = 0.1$ \label{subfig:g}]{\includegraphics[width=5.5cm]{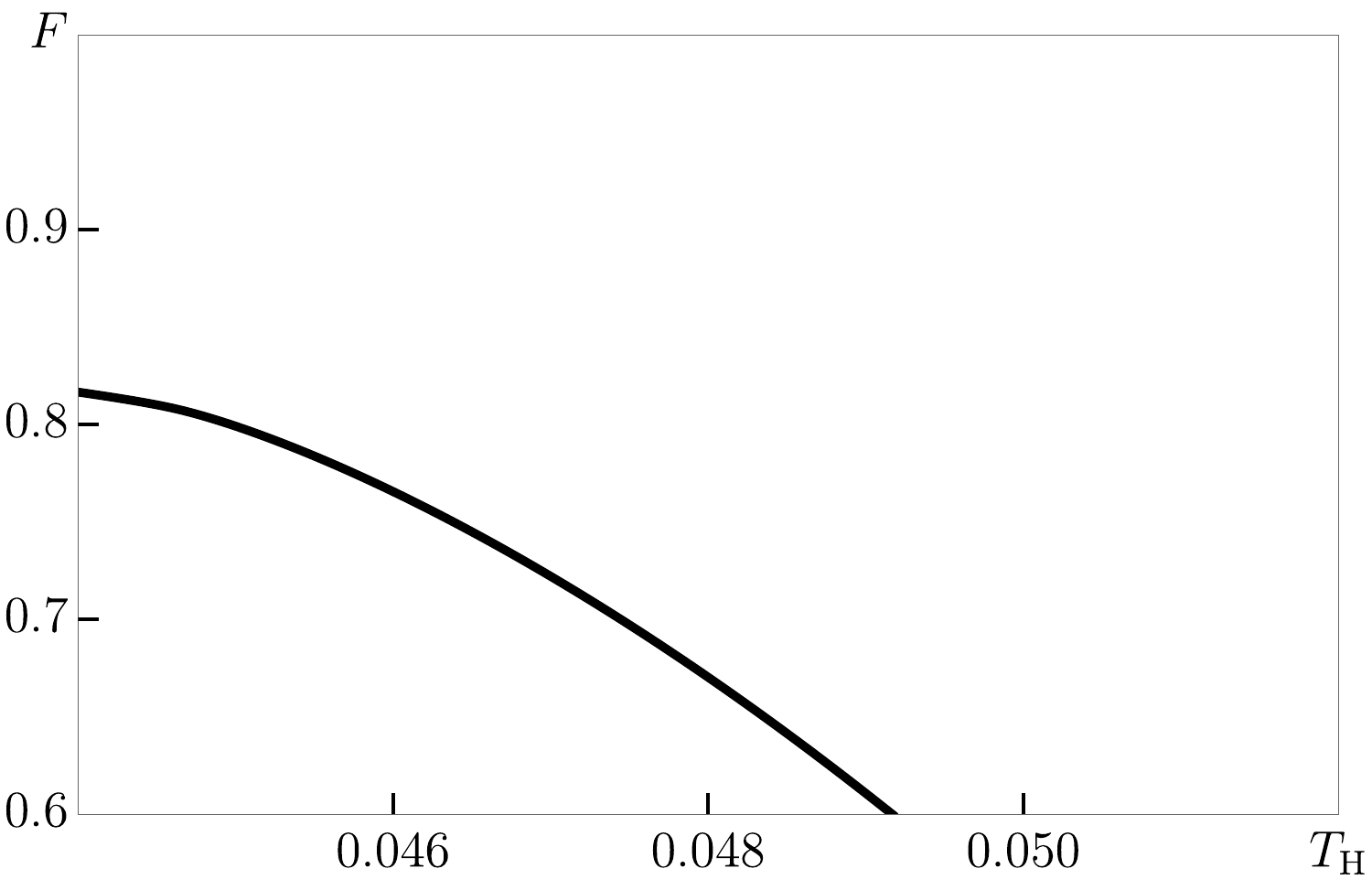}}
    \subfigure[$\Lambda = -0.09$, $\alpha = 0.5$ \label{subfig:h}]{\includegraphics[width=5.5cm]{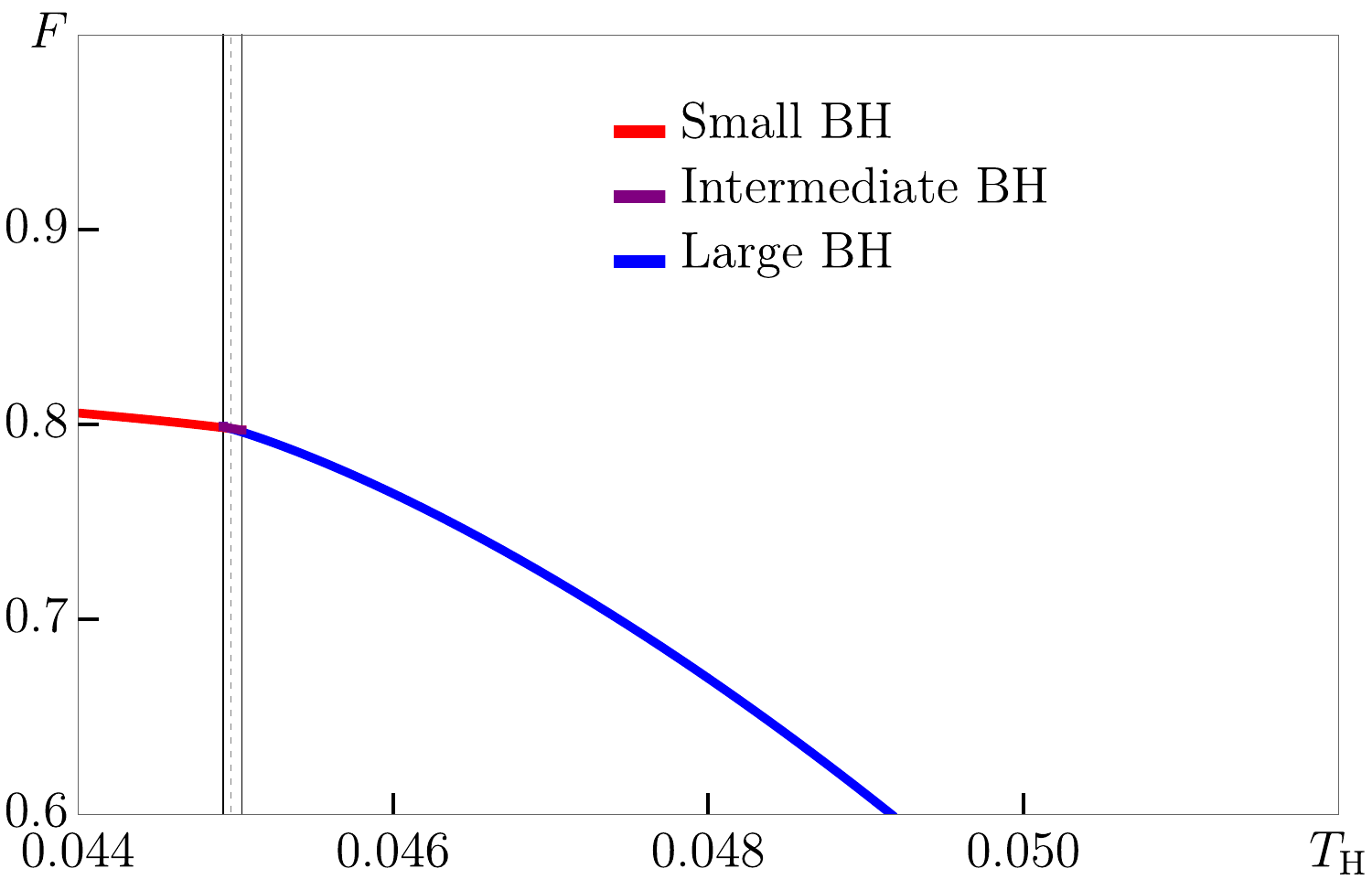}}
    \subfigure[$\Lambda = -0.09$, $\alpha = 0.9$]{\includegraphics[width=5.5cm]{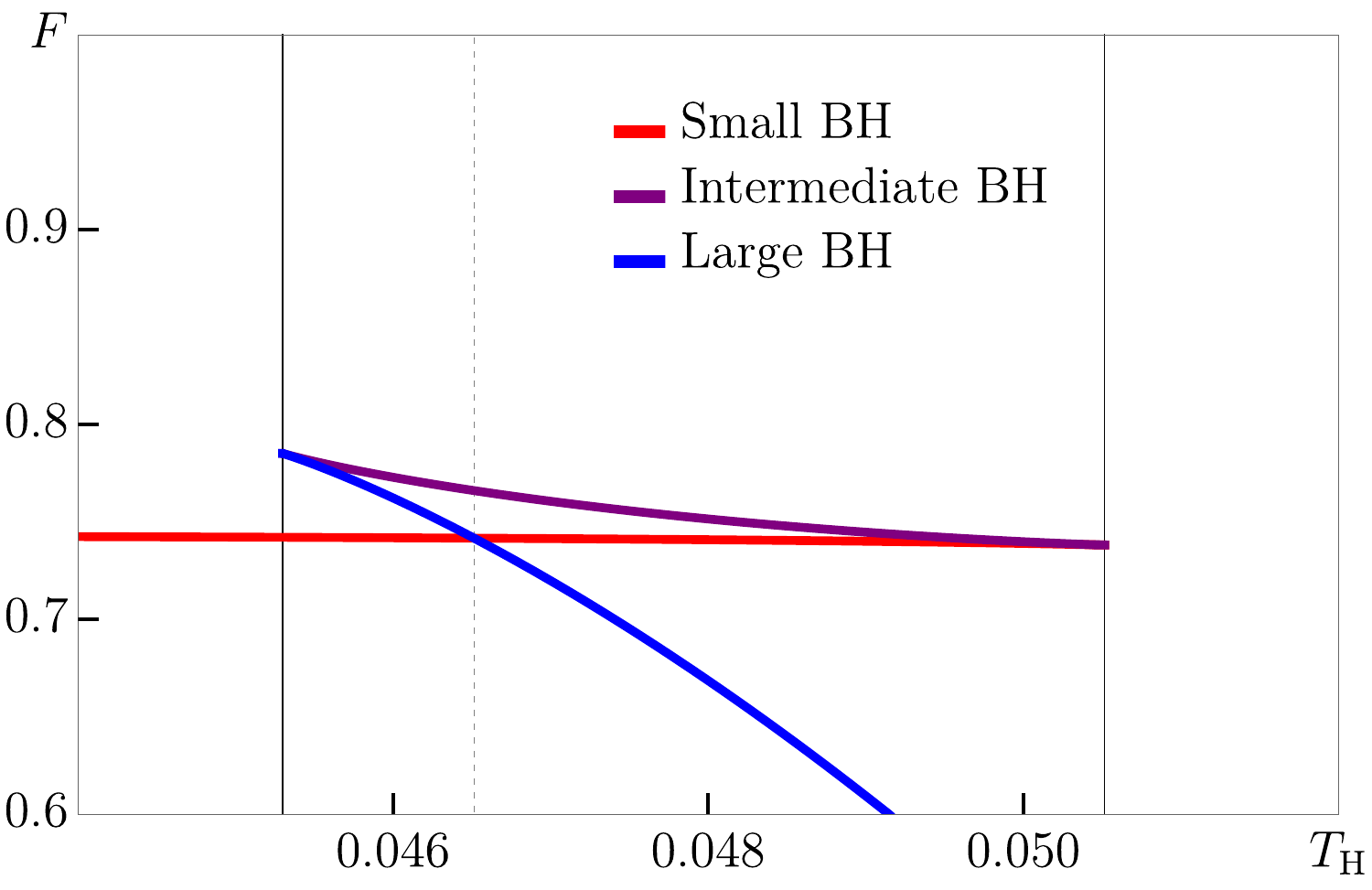}}
    \caption{Free energy as a function of Hawking temperature in the dilatonic RN--AdS black hole at different values of the cosmological constant $\Lambda$ and dilaton coupling constant $\alpha$.}
    \label{fig:phase_structure_AdS}
\end{figure}
    Fig.~\ref{fig:phase_structure_AdS} shows the free energy as a function of the Hawking temperature in asymptotically AdS spacetime for $\Lambda = -0.01$, $-0.05$, and $-0.09$, and for $\alpha = 0.1$, $0.5$, and $0.9$. Except for the case shown in Fig.~\ref{subfig:g}, the free energy exhibits a multi-valued structure within a finite temperature interval, characterized by a swallow-tail curve that reflects the coexistence of three branches corresponding to the small, intermediate, and large black hole phases, shown as red, purple, and blue lines, respectively. The vertical solid black lines indicate the turning point temperatures at which the swallow-tail structure vanishes, while the gray dashed line denotes the Hawking temperature at which the free energies of the small and large black hole branches coincide. A first-order phase transition occurs at the temperature where the two branches have equal free energy, with the thermodynamically preferred configuration determined by the global minimum of the free energy. The corresponding transition temperatures are listed in Table~\ref{tab:first_order_phase_transition}.
\begin{table}[H]
    \centering \renewcommand{\arraystretch}{1.5}
    \begin{tabular}{|c|c|c|c|c|} \hline
        & $\alpha = 0.1$ & $\alpha = 0.5$ & $\alpha = 0.9$
        \\ \hline
        $\Lambda = -0.01$ & $0.0173$ & $0.0174$ & \textbf{---}
        \\ \hline
        $\Lambda = -0.05$ & $0.0354$ & $0.0357$ & $0.0366$
        \\ \hline
        $\Lambda = -0.09$ & \textbf{---} & $0.0450$ & $0.0465$
        \\ \hline
    \end{tabular}
    \caption{Hawking temperatures of the first-order phase transition for Fig.~\ref{fig:phase_structure_AdS}.}
    \label{tab:first_order_phase_transition}
\end{table}
    For the values of $\alpha$ considered here, the transition temperature increases as $\alpha$ increases. For the case shown in Fig.~\ref{subfig:c}, the minimum mass of the black hole is not sufficiently small to allow an intersection between the free energy branches of the small and large black holes, and no first-order phase transition occurs. In Fig.~\ref{subfig:g}, the free energy remains single-valued over the entire temperature range studied, with no first-order phase transition.
    
    The first-order phase transition persists up to a critical point in the parameter space, beyond which the multi-valued free energy structure disappears. At this critical point, the swallow-tail curve collapses to a single point, at which the small and large black hole branches disappear and the free energy becomes single-valued. This critical point corresponds to the inflection point of the Hawking temperature with respect to the horizon radius, satisfying the following relation
\begin{equation}
    \left( \frac{\partial T_\mathrm{H}}{\partial r_\mathrm{h}} \right)_{Q, \;\! \alpha} = \left( \frac{\partial^2 T_\mathrm{H}}{\partial r_\mathrm{h}^2} \right)_{Q, \;\! \alpha} = 0.
\end{equation}
    In the present model, the critical parameters are obtained numerically. For the range of the dilaton coupling parameter $\alpha$ considered here, as $\alpha$ increases, the critical temperature $T_c$ increases, the critical horizon radius $r_c$ decreases, and the magnitude of the critical cosmological constant $|\Lambda_c|$ increases, indicating a systematic change in the phase structure. The corresponding numerical values are listed in Table~\ref{tab:second_order}.
\begin{table}[H]
    \centering \renewcommand{\arraystretch}{1.5}
    \begin{tabular}{|c|c|c|c|c|} \hline
        & $\alpha = 0.1$ & $\alpha = 0.5$ & $\alpha = 0.9$
        \\ \hline
        $r_c$ & $2.445$ & $2.315$ & $1.866$
        \\ \hline
        $T_c$ & $0.043$ & $0.046$ & $0.055$
        \\ \hline
        $\Lambda_c$ & $-0.084$ & $-0.094$ & $-0.148$
        \\ \hline
    \end{tabular}
    \caption{Critical parameters of the second-order phase transition for $\alpha = 0.1$, $0.5$ and $0.9$.}
    \label{tab:second_order}
\end{table}
    For Fig.~\ref{subfig:g}, the chosen value $\Lambda = -0.09$ exceeds, in magnitude, the critical value $|\Lambda_c| = 0.084$, placing the system beyond the critical regime. Consequently, the swallow-tail structure disappears and the free energy remains single-valued over the entire temperature range, indicating the absence of a phase transition. In contrast, as shown in Fig.~\ref{subfig:h}, the same value $\Lambda = -0.09$ satisfies $|\Lambda| < |\Lambda_c|$ with $\Lambda_c = -0.094$. Thus, the system remains within the phase transition regime, in which the coexistence of small and large black hole phases is still permitted. In this case, the swallow-tail structure is still present, although it is considerably smaller, indicating that the first-order phase transition becomes weaker as the critical point is approached.

\begin{figure}[H]
    \centering
    \subfigure[$\Lambda = -0.01$, $\alpha = 0.1$]{\includegraphics[width=5.5cm]{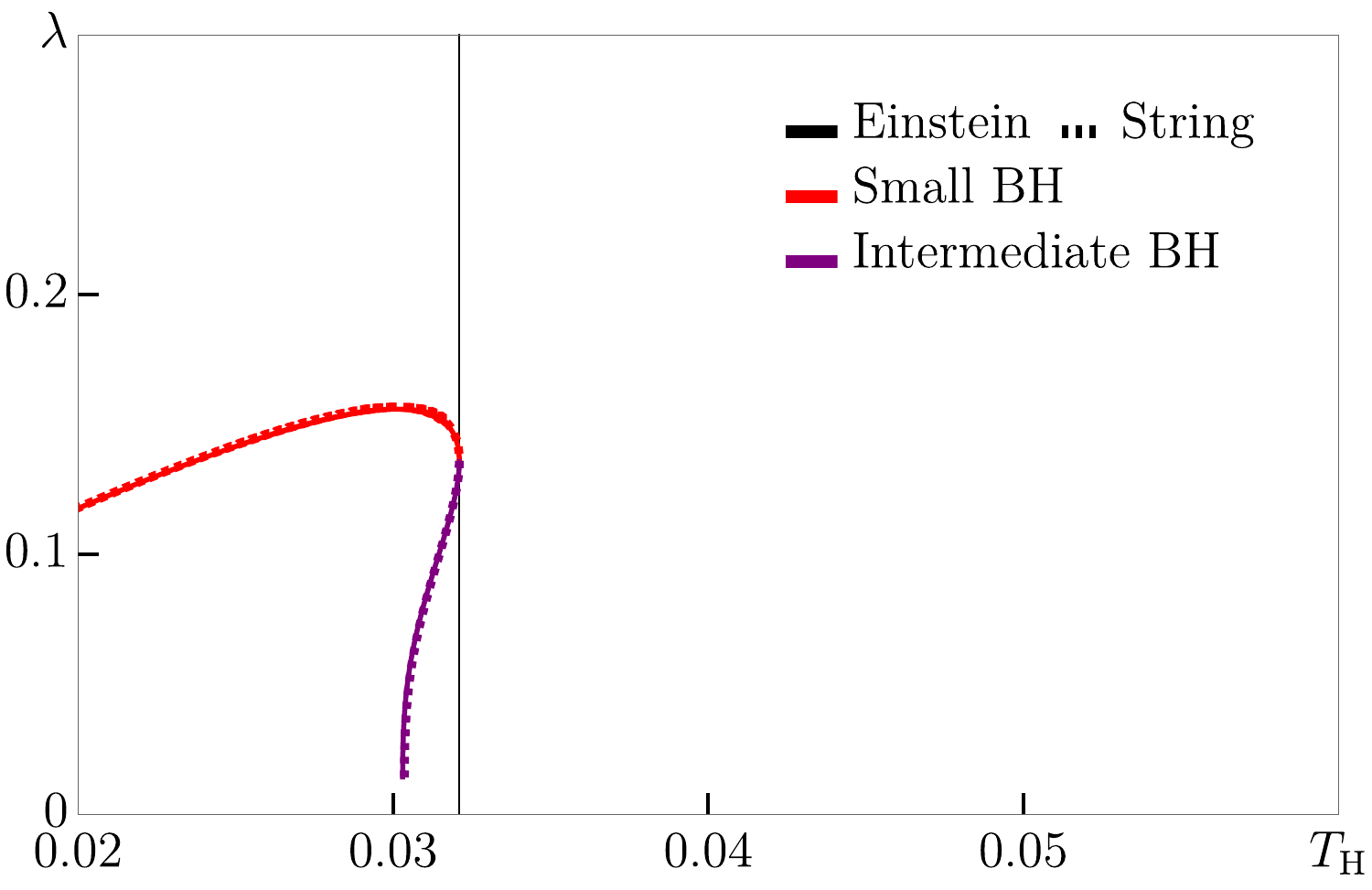}}
    \subfigure[$\Lambda = -0.01$, $\alpha = 0.5$]{\includegraphics[width=5.5cm]{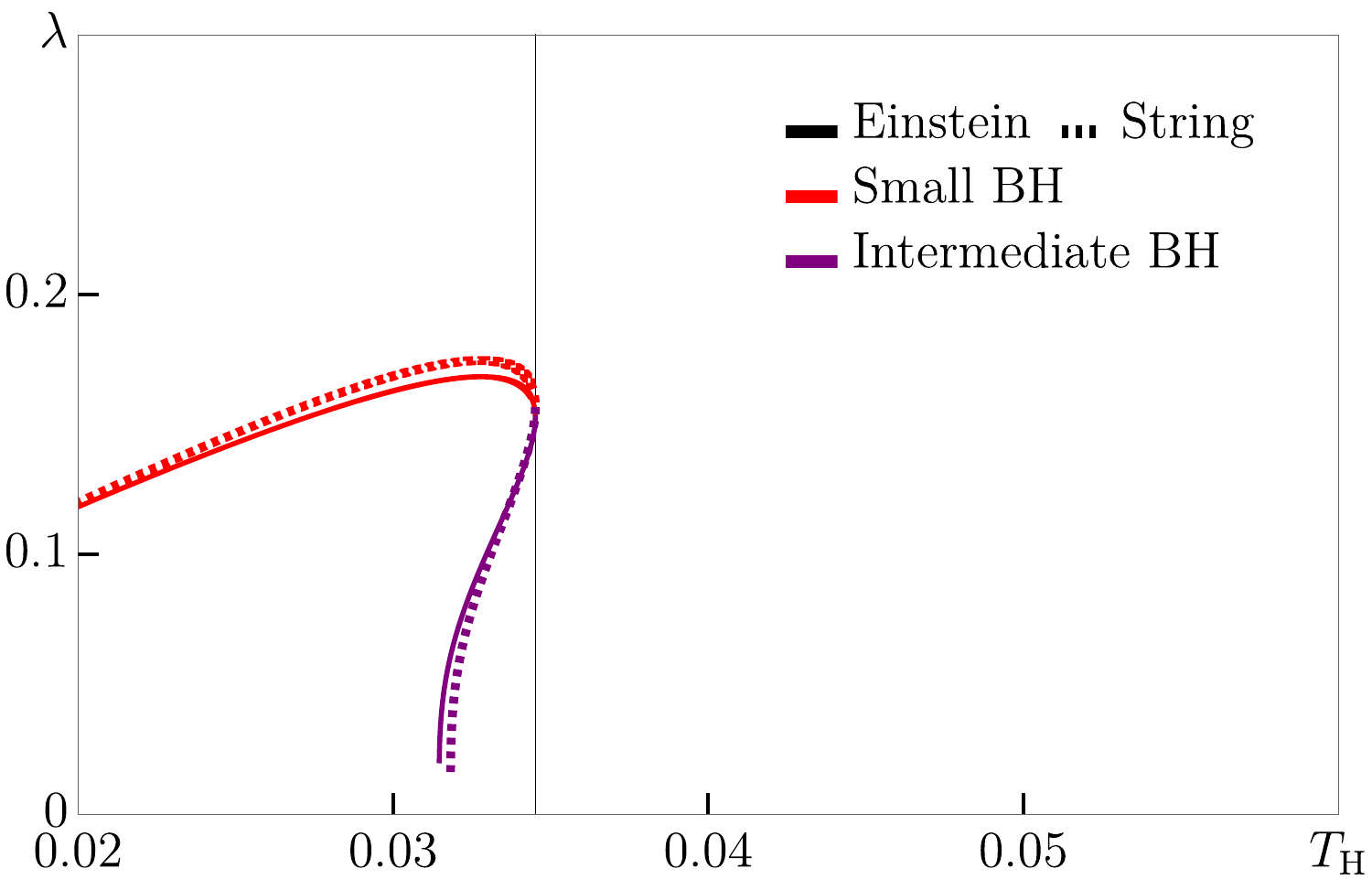}}
    \subfigure[$\Lambda = -0.01$, $\alpha = 0.9$]{\includegraphics[width=5.5cm]{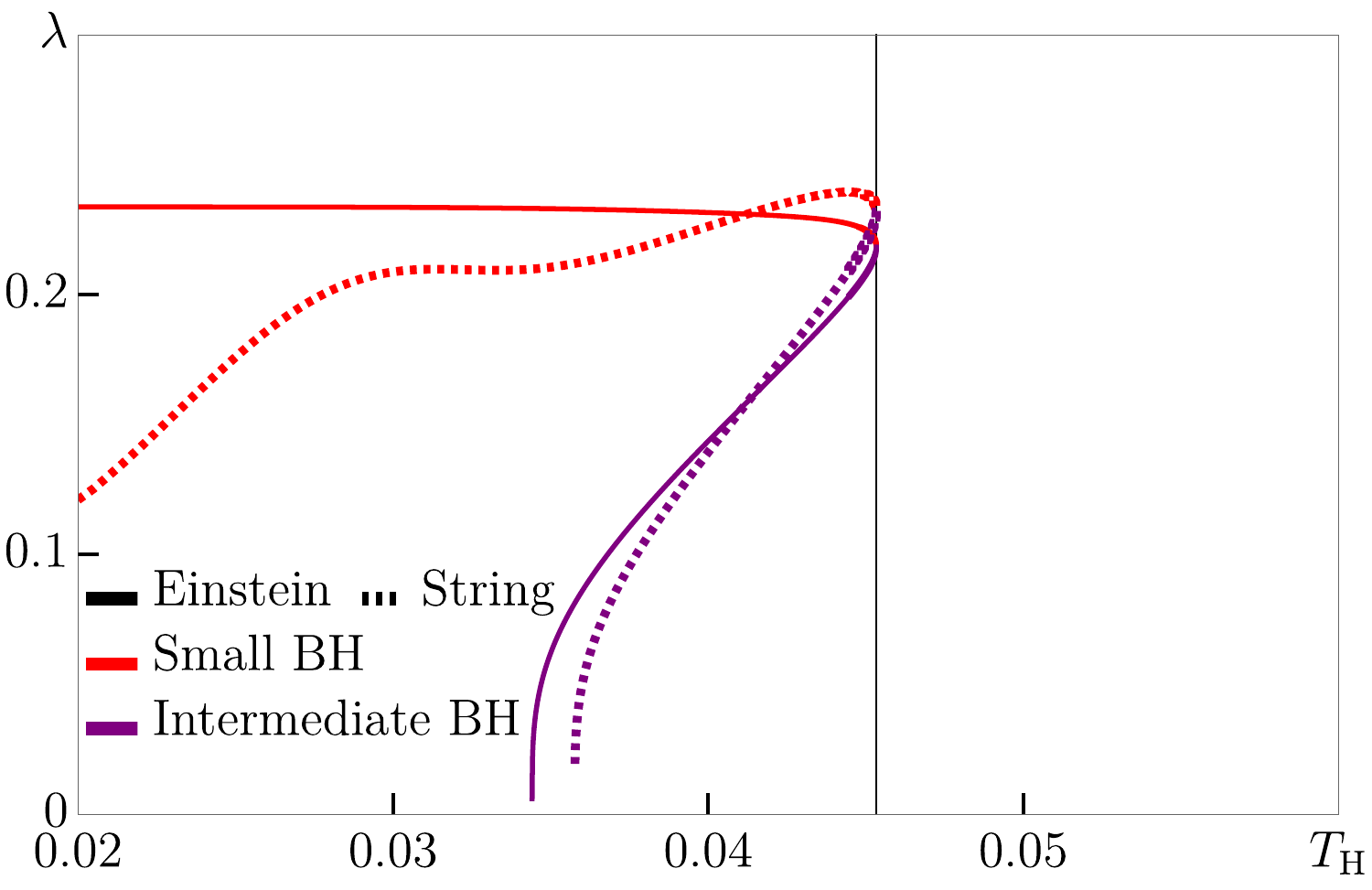}}
    \subfigure[$\Lambda = -0.05$, $\alpha = 0.1$ \label{subfig:Lyapunov_exponent_AdS_d}]{\includegraphics[width=5.5cm]{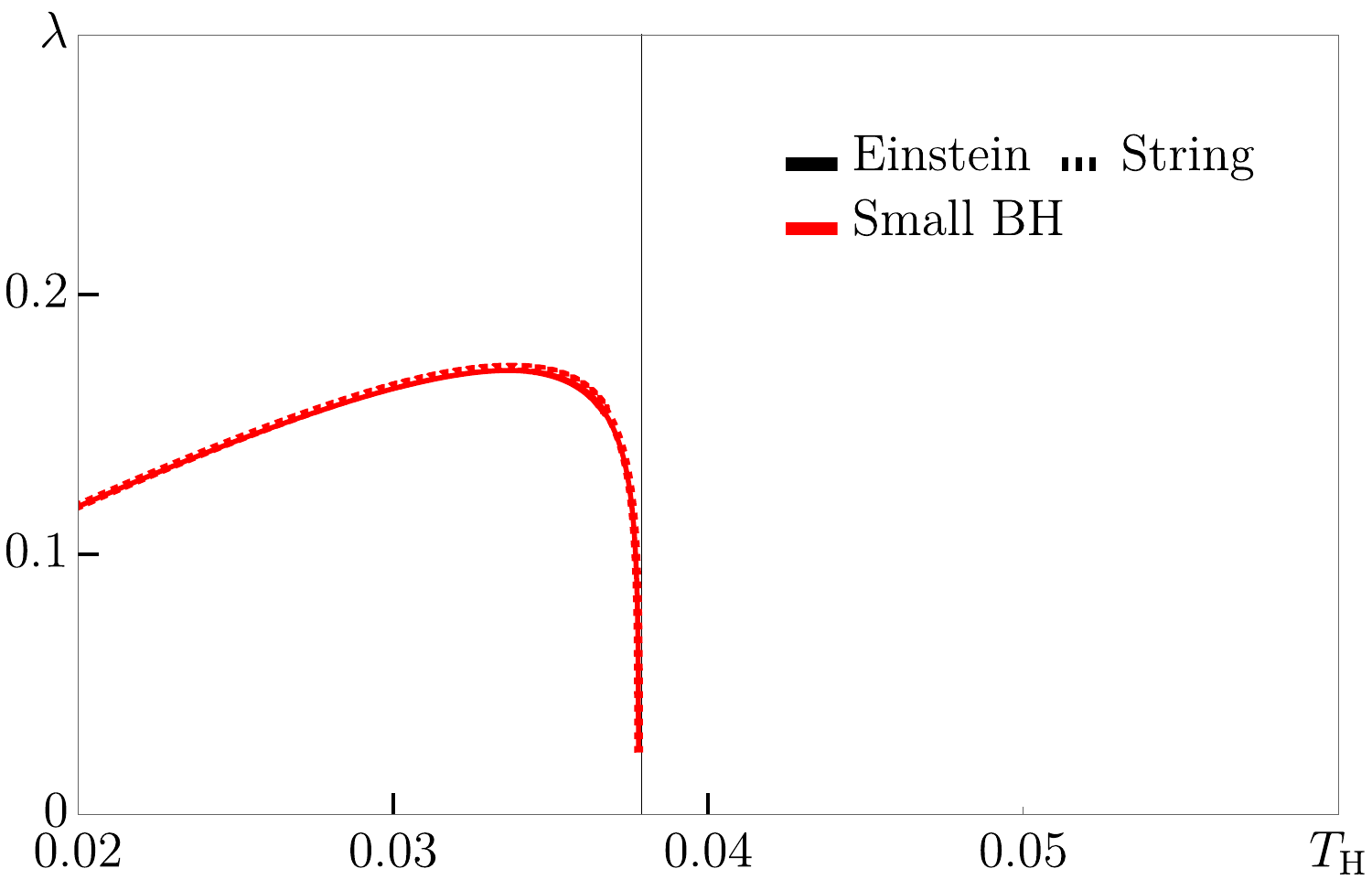}}
    \subfigure[$\Lambda = -0.05$, $\alpha = 0.5$  \label{subfig:Lyapunov_exponent_AdS_e}]{\includegraphics[width=5.5cm]{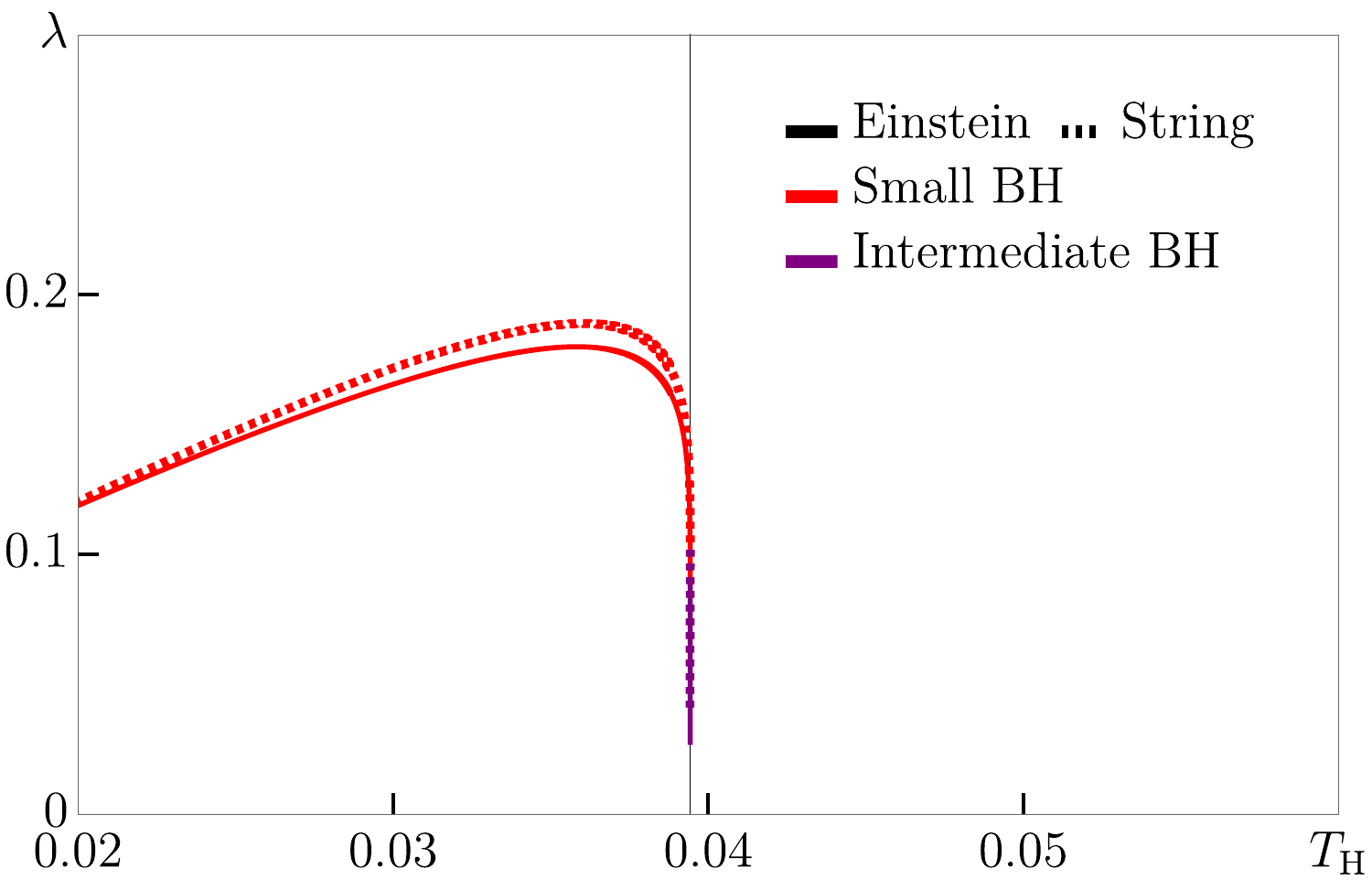}}
    \subfigure[$\Lambda = -0.05$, $\alpha = 0.9$  \label{subfig:Lyapunov_exponent_AdS_f}]{\includegraphics[width=5.5cm]{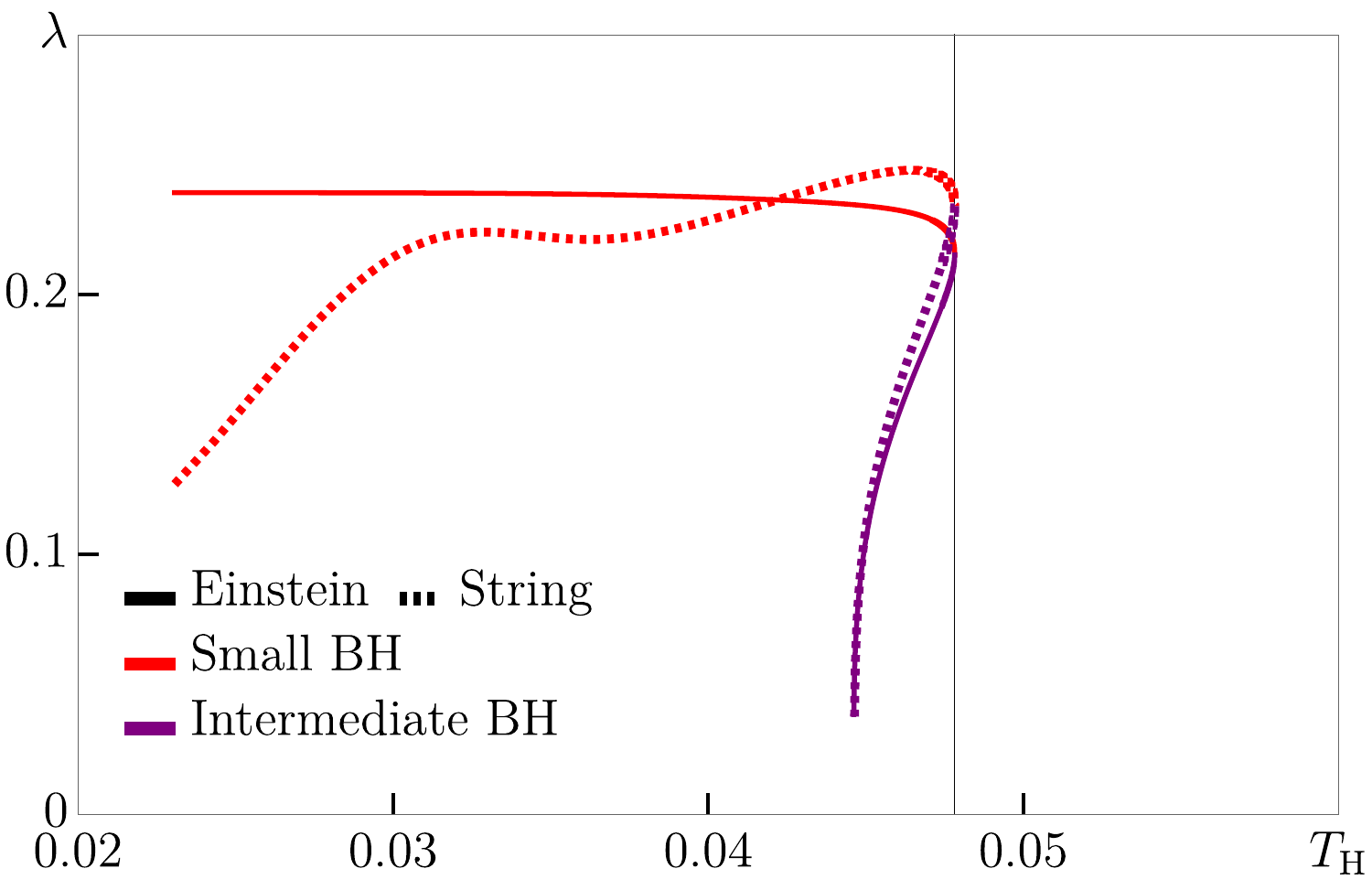}}
    \subfigure[$\Lambda = -0.09$, $\alpha = 0.1$ \label{subfig:Lyapunov_exponent_AdS_g}]{\includegraphics[width=5.5cm]{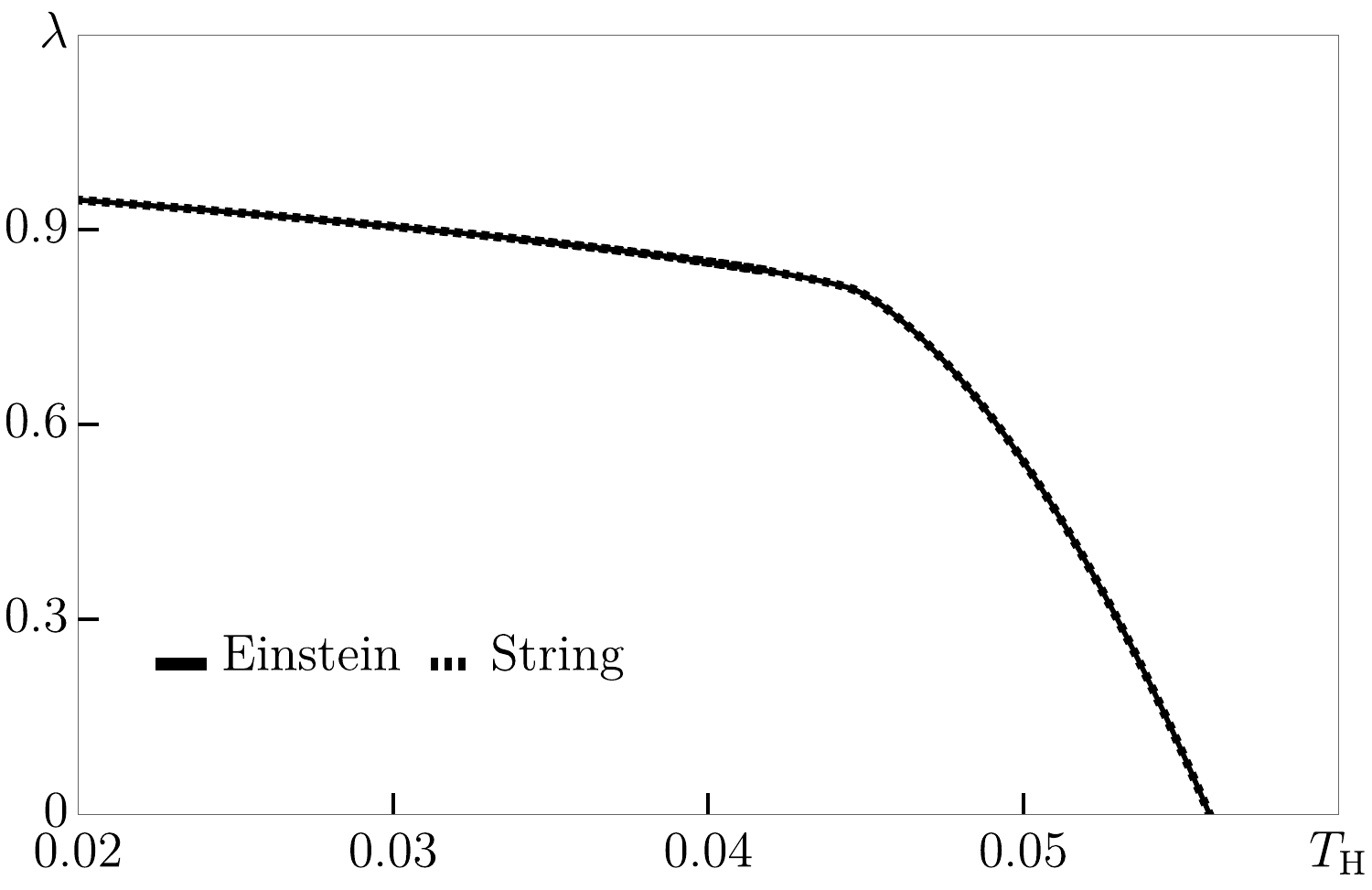}}
    \subfigure[$\Lambda = -0.09$, $\alpha = 0.5$ \label{subfig:Lyapunov_exponent_AdS_h}]{\includegraphics[width=5.5cm]{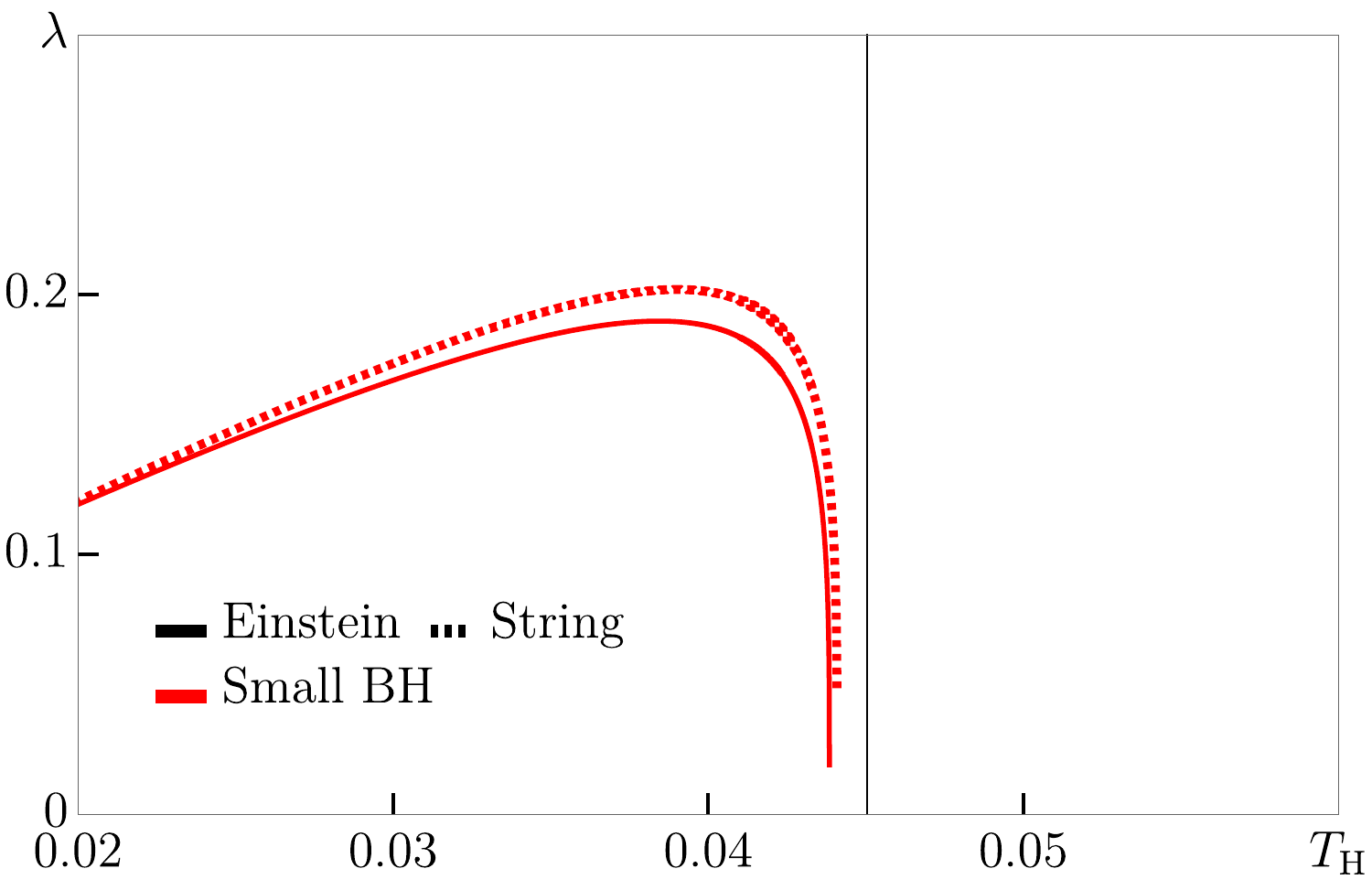}}
    \subfigure[$\Lambda = -0.09$, $\alpha = 0.9$]{\includegraphics[width=5.5cm]{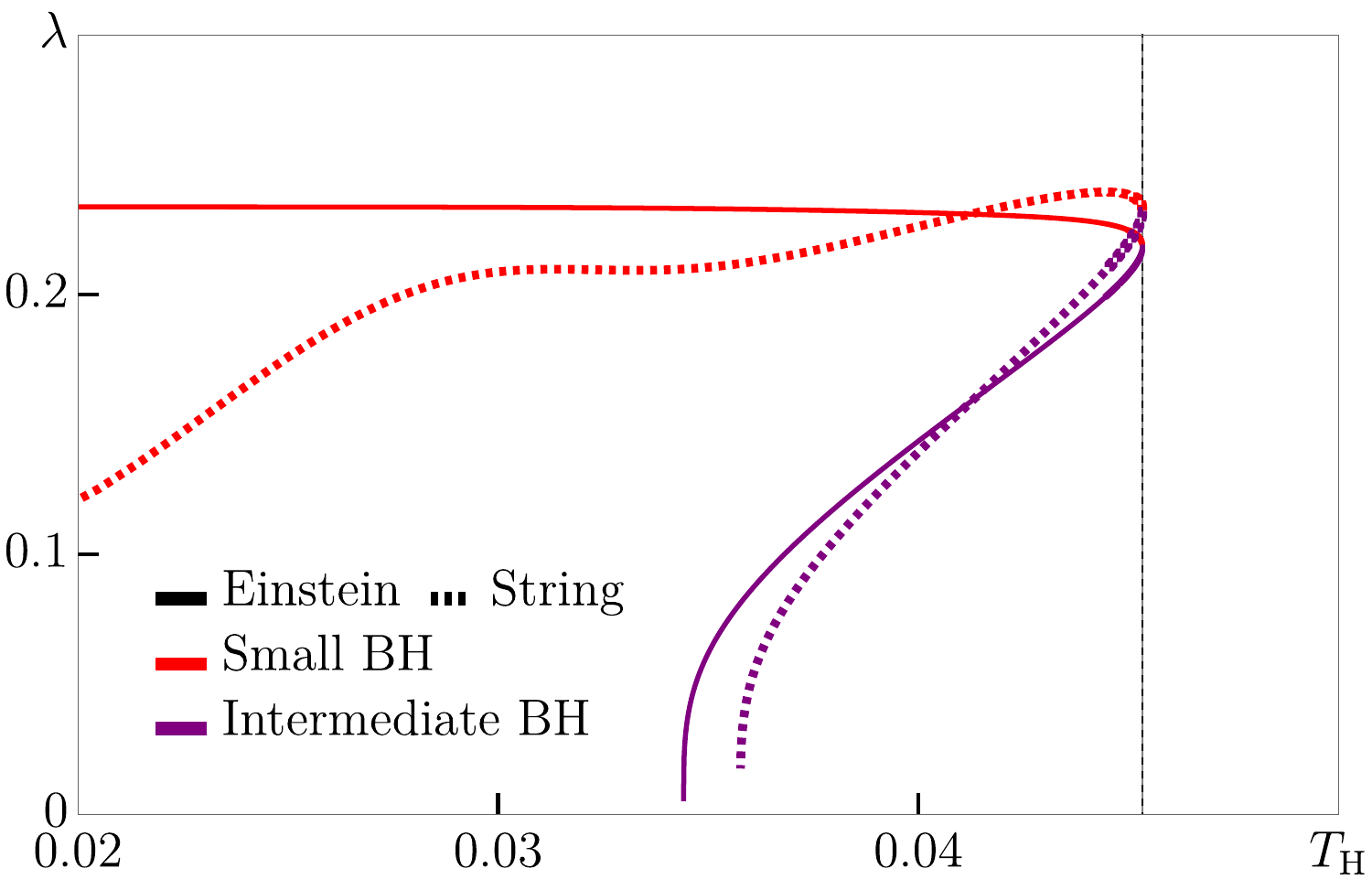}}
    \caption{Lyapunov exponents as a function of the Hawking temperature for the dilatonic RN--AdS black hole with $\alpha = 0.1$, $0.5$, $0.9$, where $q = 2$ and $L = 0.01$.}
    \label{fig:Lyapunov_exponent_AdS}
\end{figure}
    Fig.~\ref{fig:Lyapunov_exponent_AdS} demonstrates the Lyapunov exponent $\lambda$ as a function of the Hawking temperature $T_\mathrm{H}$ for the dilatonic RN--AdS black hole, with fixed particle parameters $q = 2$ and $L = 0.01$, for several different values of the cosmological constant $\Lambda$ and the dilaton coupling parameter $\alpha$. In all cases studied, the Einstein and string frames exhibit qualitatively similar thermodynamic branch structures, while the magnitude of the Lyapunov exponent shows explicit frame dependence induced by dilaton coupling. For a small value of $\alpha$ $(= 0.1)$, the Lyapunov exponents in the Einstein and string frames are nearly indistinguishable over the entire allowed temperature range. As $\alpha$ increases, the discrepancy between the two frames increases progressively, and for $\alpha = 0.9$, the separation between the solid and dashed curves becomes pronounced. The temperature dependence of the Lyapunov exponent captures the thermodynamic phase structure identified by the free energy, as shown in Fig.~\ref{fig:phase_structure_AdS}. In parameter regions where the free energy exhibits a swallow-tail structure, the Lyapunov exponent displays a cusp, reflecting the coexistence of black hole branches. In contrast to the free energy, which captures two distinct transitions among the small, intermediate, and large black hole branches, the Lyapunov exponent exhibits a single transition in the same region. The limited correspondence in this region is attributed to the absence of unstable circular orbits within the large black hole branch for the chosen values of $q$ and $L$. Consequently, the dynamical probe is restricted to the small and intermediate black hole branches, while the large black hole branch remains beyond the scope of the dynamical analysis. More specifically, in Figs.~\ref{subfig:Lyapunov_exponent_AdS_d} and~\ref{subfig:Lyapunov_exponent_AdS_h}, unstable circular orbits exist exclusively within the small black hole branch. The absence of unstable circular orbits in the large black hole regime implies that gravitational attraction is dominant compared with electromagnetic repulsion and centrifugal effects for small particle charges and angular momentum. For sufficiently large charges or angular momentum, particles can form unstable circular orbits even in the large black hole regime. Computing the Lyapunov exponent for these orbits reproduces the cusp-like structure corresponding to the transition between intermediate and large black holes. In Fig.~\ref{subfig:Lyapunov_exponent_AdS_g}, the parameter region beyond the critical values is shown, in which the swallow-tail structure vanishes and the free energy becomes single-valued. Consequently, the Lyapunov exponent also becomes single-valued, eliminating the cusp-like structure at which its derivative diverges.
    
\begin{figure}[H]
    \centering
    \subfigure[$\lambda(r_\mathrm{h})$ for $\Lambda = -0.05$, $\alpha = 0.5$ \label{subfig:difference_AdS_a}]{\includegraphics[width=5.5cm]{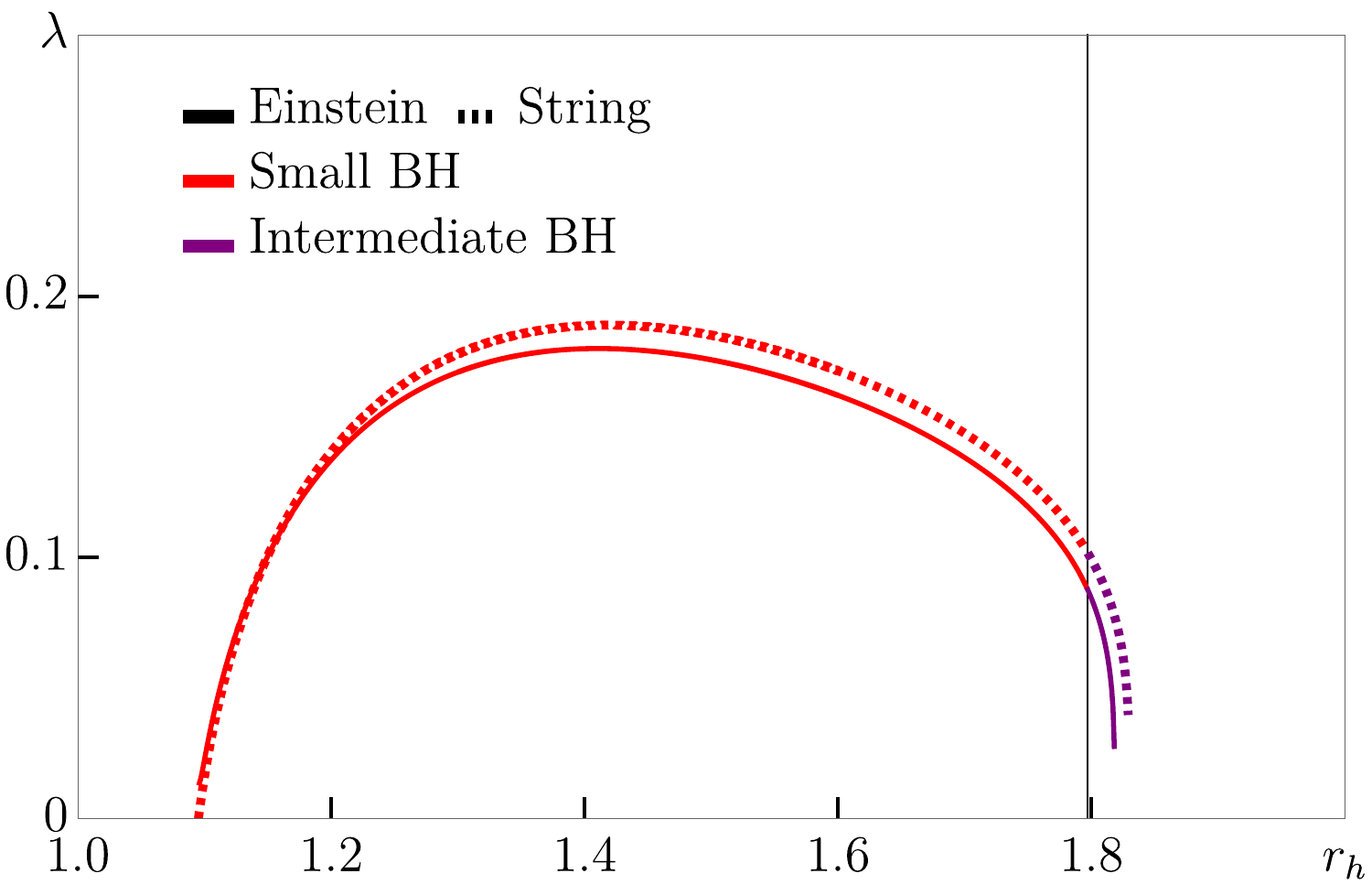}}
    \subfigure[$\Delta\lambda(r_\mathrm{h})$ for $\Lambda = -0.05$, $\alpha = 0.5$ \label{subfig:difference_AdS_b}]{\includegraphics[width=5.5cm]{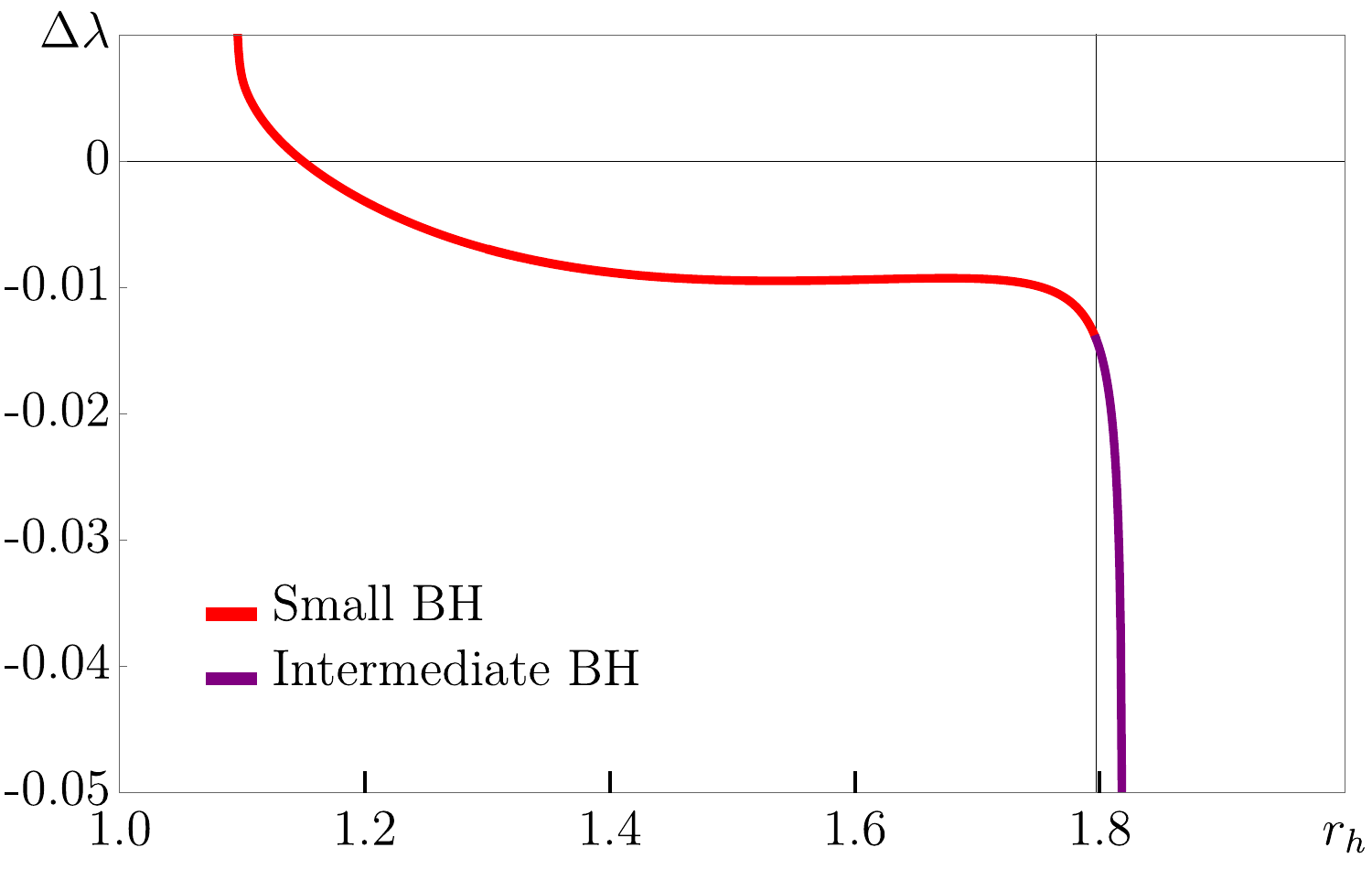}}
    \subfigure[$\Delta\lambda(T_\mathrm{H})$ for $\Lambda = -0.05$, $\alpha = 0.5$ \label{subfig:difference_AdS_c}]{\includegraphics[width=5.5cm]{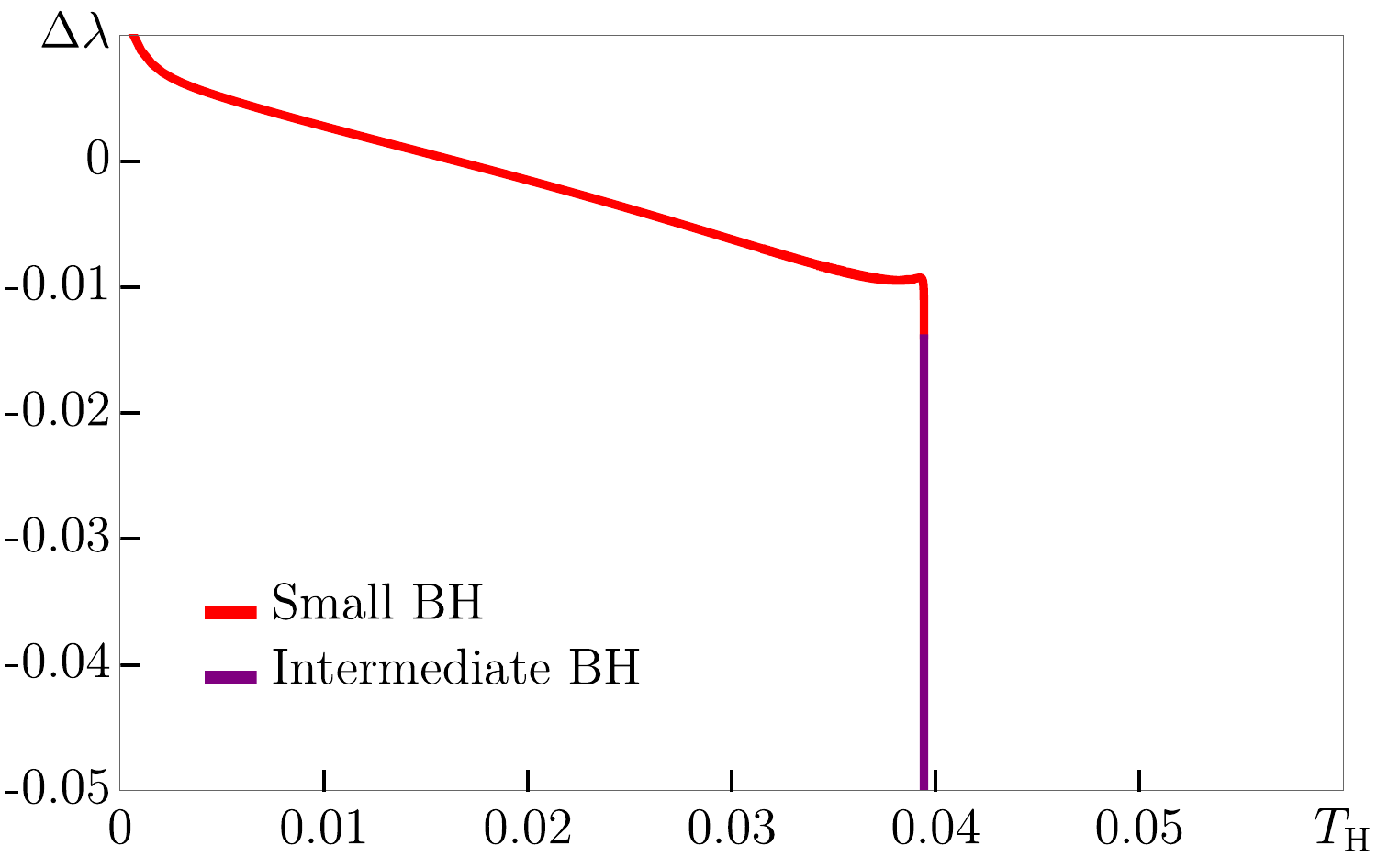}}
    \subfigure[$\lambda(r_\mathrm{h})$ for $\Lambda = -0.05$, $\alpha = 0.9$ \label{subfig:difference_AdS_d}]{\includegraphics[width=5.5cm]{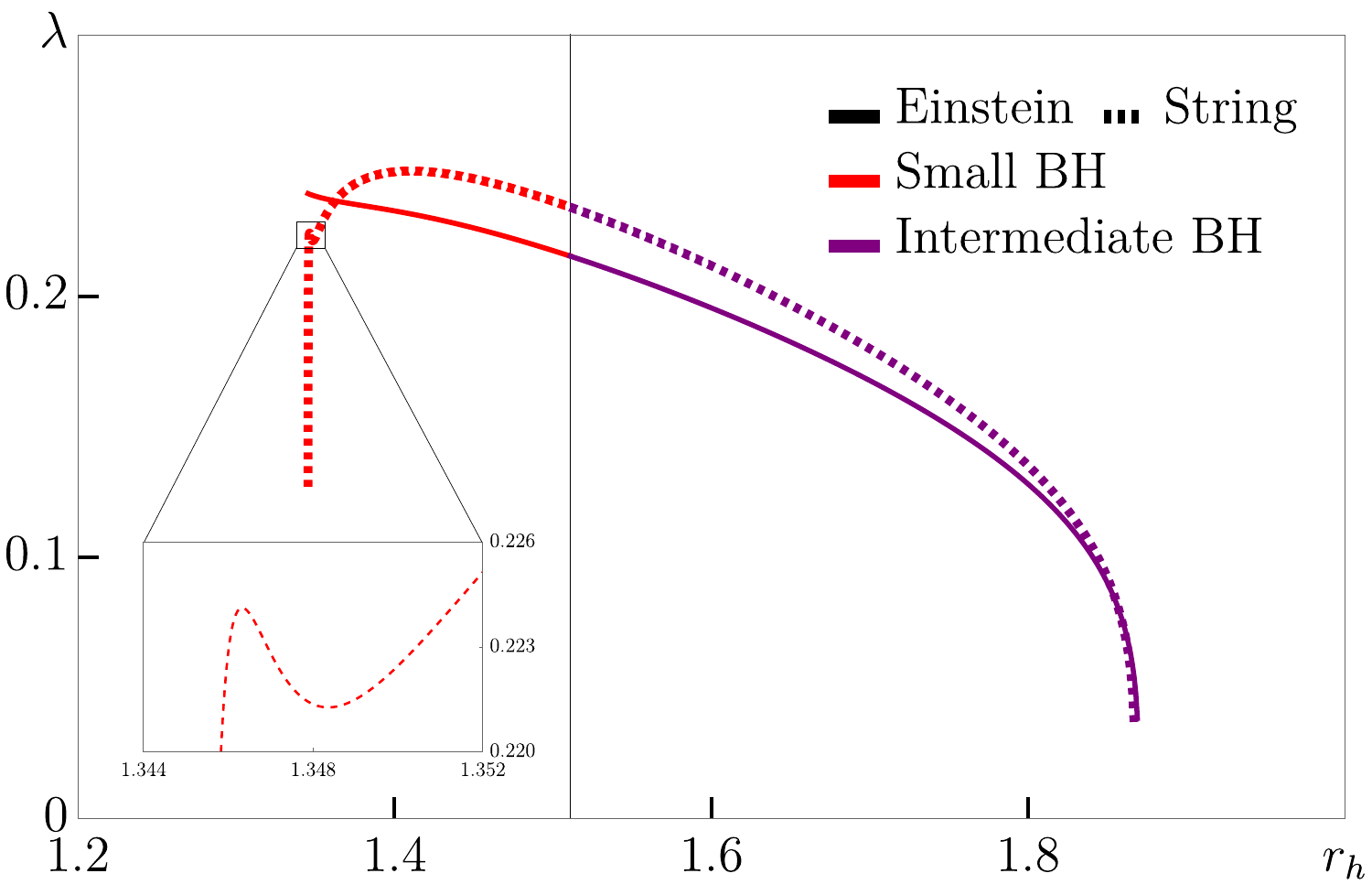}}
    \subfigure[$\Delta\lambda(r_\mathrm{h})$ for $\Lambda = -0.05$, $\alpha = 0.9$ \label{subfig:difference_AdS_e}]{\includegraphics[width=5.5cm]{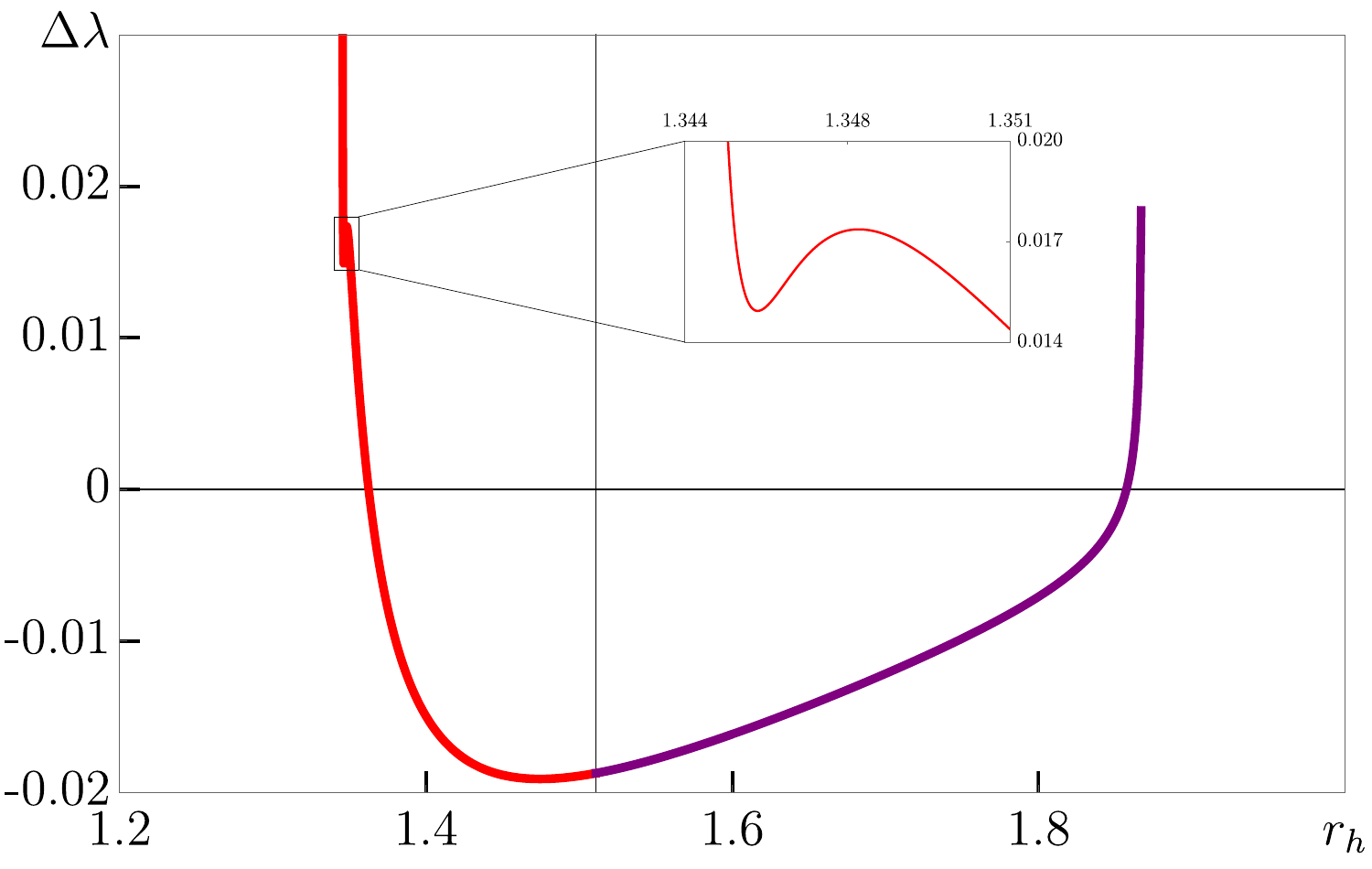}}
    \subfigure[$\Delta\lambda(T_\mathrm{H})$ for $\Lambda = -0.05$, $\alpha = 0.9$ \label{subfig:difference_AdS_f}]{\includegraphics[width=5.5cm]{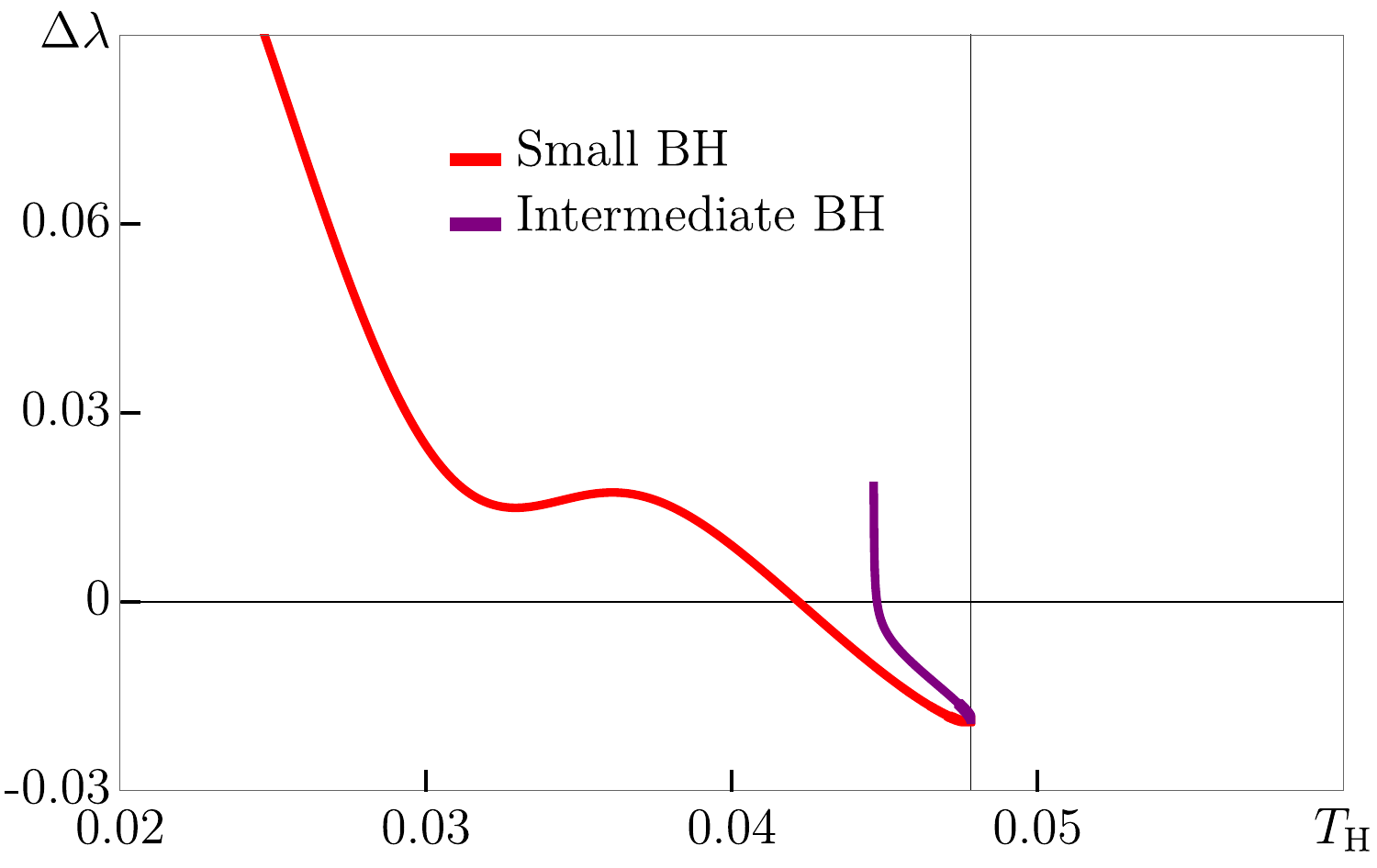}}
    \caption{Lyapunov exponents and their difference in the Einstein and string frames.}
    \label{fig:difference_AdS}
\end{figure}
    Fig.~\ref{fig:difference_AdS} depicts a detailed comparison of the Lyapunov exponent $\lambda$ and its frame difference $\Delta\lambda \equiv \lambda_\mathrm{E} - \lambda_\mathrm{S}$ for Figs.~\ref{subfig:Lyapunov_exponent_AdS_e}~$(\alpha = 0.5)$ and~\ref{subfig:Lyapunov_exponent_AdS_f}~$(\alpha = 0.9)$, with $\Lambda = -0.05$. From left to right, the columns display $\lambda(r_\mathrm{h})$, $\Delta\lambda(r_\mathrm{h})$, and $\Delta\lambda(T_\mathrm{H})$, respectively. As shown in Fig.~\ref{subfig:difference_AdS_a}, the Lyapunov exponent in both the Einstein and string frames initially increases, reaches a maximum, and subsequently decreases over the range of $r_\mathrm{h}$ values tested. Figs.~\ref{subfig:difference_AdS_b}~and~\ref{subfig:difference_AdS_c} show that the Lyapunov exponent in the Einstein frame is larger in the earliest part of the small black hole regime. Beyond this region, the Lyapunov exponent in the string frame is larger, yielding negative values of $\Delta\lambda$. As shown in Fig.~\ref{subfig:difference_AdS_d}, as $r_\mathrm{h}$ increases, the Einstein and string frames show monotonic and non-monotonic behavior, respectively. In the small black hole regime, as $r_\mathrm{h}$ increases, $\lambda_\mathrm{S}$ initially increases, then decreases over a very narrow interval, and subsequently increases again before reaching a local maximum and eventually decreasing. These dynamics are further elucidated in Fig.~\ref{subfig:difference_AdS_e}, which illustrates the difference between the Lyapunov exponents, $\Delta\lambda(r_\mathrm{h})$. In the small-$r_\mathrm{h}$ regime, as $\lambda_\mathrm{S}$ increases, $\Delta\lambda$ is initially large and then decreases. Over the narrow interval where $\lambda_\mathrm{S}$ decreases, $\Delta\lambda$ increases correspondingly and subsequently decreases until $\lambda_\mathrm{S}$ attains a local maximum. As $r_\mathrm{h}$ increases further, the Lyapunov exponent in the Einstein frame becomes dominant, yielding positive values of $\Delta\lambda$. In Fig.~\ref{subfig:difference_AdS_f}, the Lyapunov exponent in the Einstein frame remains dominant over most of the Hawking temperature range, while that in the string frame exceeds it only within a narrow transition region between the small and intermediate black hole branches.

\begin{figure}[H]
    \centering
    \subfigure[Lyapunov exponent as a function of $r_\mathrm{h}$ \label{subfig:Lyapunov_exponent_AdS_large_q_a}]{\includegraphics[width=8.2cm]{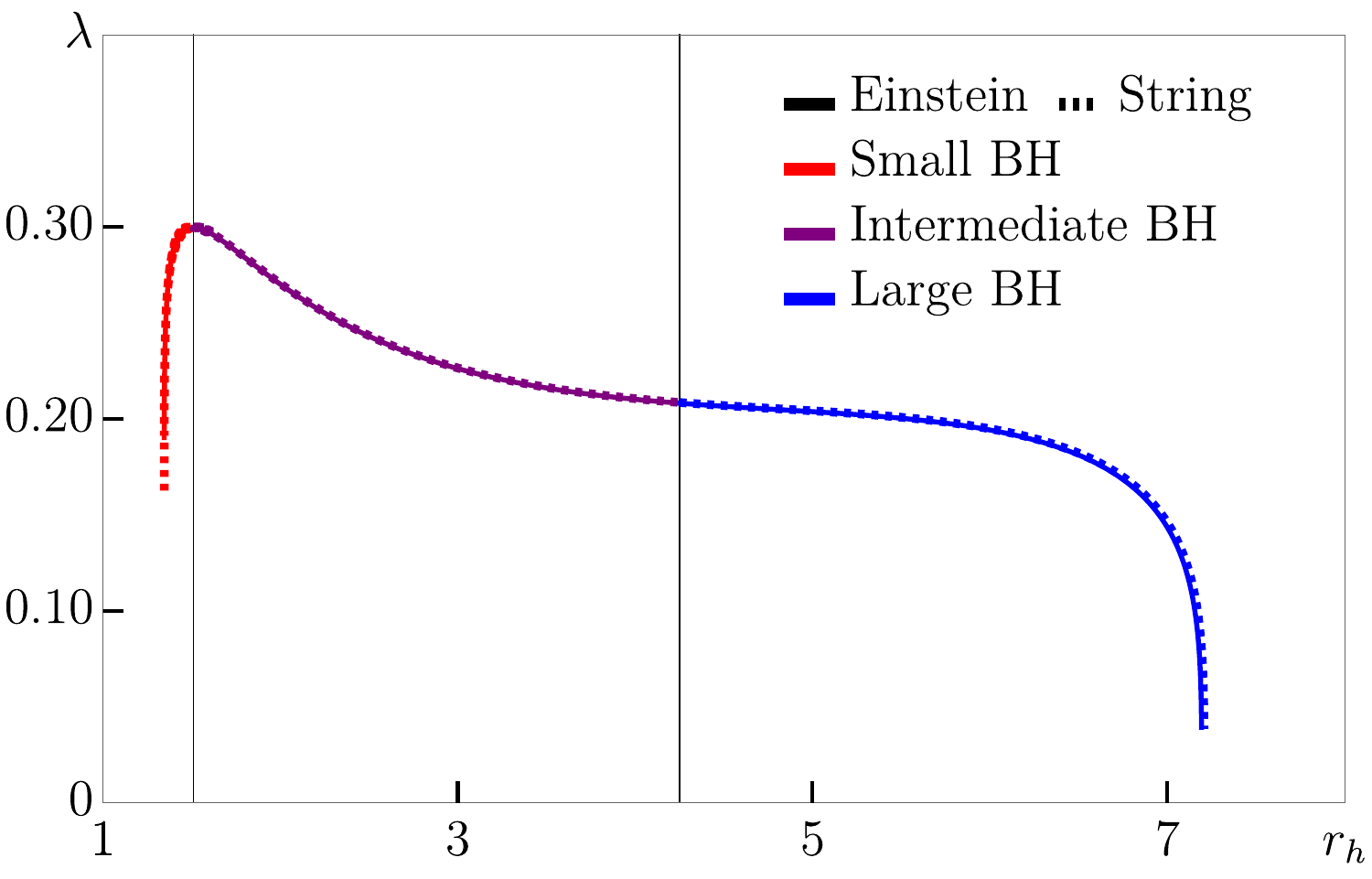}}
    \subfigure[Lyapunov exponent as a function of $T_\mathrm{H}$ \label{subfig:Lyapunov_exponent_AdS_large_q_b}]{\includegraphics[width=8.2cm]{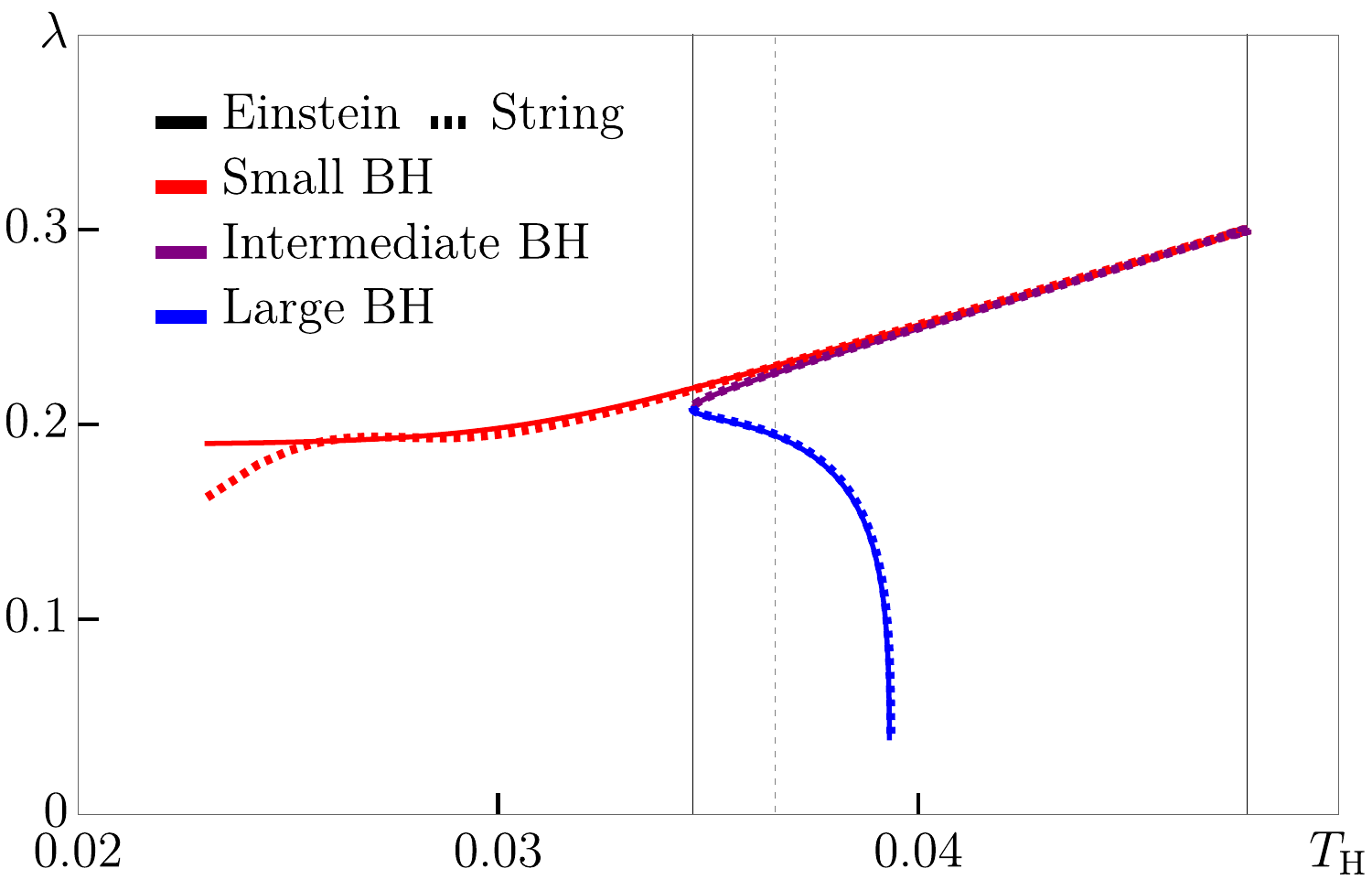}}
    \caption{Lyapunov exponents for the dilatonic RN--AdS black hole with $\Lambda = -0.05$ and $\alpha = 0.5$, where $q = 20$ and $L = 0.01$.}
    \label{fig:Lyapunov_exponent_AdS_large_q}
\end{figure}
    Fig.~\ref{fig:Lyapunov_exponent_AdS_large_q} displays the Lyapunov exponent for $q = 20$ and $L = 0.01$, corresponding to a larger particle charge than that considered in Fig.~\ref{fig:Lyapunov_exponent_AdS}. In the configurations shown in Fig.~\ref{fig:Lyapunov_exponent_AdS}, circular orbits are confined to the small and intermediate black hole branches, with no circular orbits in the large black hole regime. This behavior is attributed to the relatively weak electromagnetic repulsion associated with a smaller particle charge, which is insufficient to balance the gravitational attraction in the large black hole regime. For the larger value of $q$ considered here, the increased repulsive interaction allows circular orbits to persist even for large black holes, thereby enabling the computation of the corresponding Lyapunov exponent in the large black hole regime.

\section{Conclusion} \label{sec:conclusion}
    In this study, the frame dependence of the thermodynamic quantities and Lyapunov exponents of probe particles in dilatonic RN--AdS black holes was investigated within Einstein--Maxwell--dilaton theory. Recent studies have demonstrated that the Lyapunov exponent for unstable circular orbits provides a probe of black hole phase transitions based on characteristic cusp-like structures. Motivated by the relationship between the Lyapunov exponent and the thermodynamic phase structure, we examined whether the phase transition features captured by the Lyapunov exponent are preserved across the Einstein and string frames, despite the explicit frame dependence induced by the dilaton field. To this end, both the thermodynamic properties of the black hole and the dynamics of circular orbits characterized by the Lyapunov exponent were analyzed.
    
    First, the thermodynamic quantities in both frames were examined. Although the naive Bekenstein--Hawking entropy in the string frame differed from that in the Einstein frame due to the non-vanishing dilaton field at the horizon, the Wald entropy formalism yielded identical entropies in the two frames. Moreover, after an appropriate constant shift of the asymptotic value of the dilaton field, the Hawking temperatures coincided exactly. Consequently, the first law of black hole thermodynamics and the free energy were expressed by identical equations in both frames, indicating that the thermodynamic phase structure was independent of the choice of frame when the physical quantities were properly defined.

    Next, the effective dynamics of probe particles were derived, and the corresponding Lyapunov exponents in the Einstein and string frames were obtained. The resulting expressions revealed that the difference between the two frames originated from the dilaton-dependent contribution to the effective mass term of the probe particle. In the absence of the dilaton-dependent conformal factor, the descriptions in the Einstein and string frames reduced to trivially equivalent forms, yielding identical Lyapunov exponents. In the massless particle limit, the dilaton coupling vanished with the particle mass, and the Lyapunov exponent consequently remained constant under conformal transformations. In contrast, for massive particles, the dilaton field modified the particle dynamics in the string frame, leading to the explicit frame dependence of the Lyapunov exponent.

    The numerical analysis showed that the Lyapunov exponent captured the thermodynamic phase structure of the dilatonic black hole. In the asymptotically flat case, the free energy exhibited a cusp structure where the small and large black hole branches coexisted at the maximum temperature. Correspondingly, the Lyapunov exponent displayed a similar cusp-like structure, indicating a direct relation between the orbital dynamics and the thermodynamic phase structure. Although the cusp-like structure of the free energy was identical in both frames, the temperature dependence of the Lyapunov exponent differed between the Einstein and string frames. This discrepancy became more pronounced for larger values of the dilaton coupling parameter $\alpha$, for which the dilaton contribution more strongly modified the dynamics of a massive probe particle.

    In asymptotically AdS spacetime, the free energy developed a swallow-tail structure characteristic of a first-order phase transition between black hole branches. The thermodynamic phase behavior, represented by the free energy, varied systematically with the cosmological constant $\Lambda$ and the dilaton coupling parameter $\alpha$, while remaining identical in the two frames. However, a difference was observed between the thermodynamic phase structure and the structure captured by the Lyapunov exponent for certain parameter choices. In these cases, the free energy exhibited two cusps associated with the swallow-tail structure, while the Lyapunov exponent developed a single cusp. This limited correspondence was attributed to unstable circular orbits confined to the small and intermediate black hole branches for the chosen particle parameters, while they were absent in the large black hole branch due to dominant gravitational attraction.

    A detailed comparison further revealed that the dilaton coupling parameter $\alpha$ introduced non-trivial frame-dependent features in the Lyapunov exponent profiles. Depending on the value of $\alpha$, the Einstein and string frames exhibited qualitatively distinct evolutions, in which monotonic behavior in one frame contrasted with non-monotonic behavior in the other. This structural difference could reverse the relative ordering of the frames for which the Lyapunov exponent was dominant within localized regimes of black hole size and temperature. For a small particle charge $q$ and angular momentum $L$, the dilaton-dependent terms dominated because of the coupling of the dilaton field to the particle mass, thereby enhancing the discrepancy between the Einstein and string frames. In the large-$q$ regime, the enhanced electromagnetic repulsion restored unstable circular orbits in the large black hole branch and reproduced the multi-cusp structure corresponding to the swallow-tail structure of the free energy. However, over the same parameter range, the dominant electromagnetic contribution suppressed the relative effect of the mass-coupled dilaton term, leading to close agreement between the Einstein and string frames. This behavior was observed consistently in both asymptotically flat and asymptotically AdS spacetimes.

    Therefore, the analysis revealed that, for massive probe particles, the Lyapunov exponent captured the thermodynamic phase structure of the black hole and exhibited explicit dependence on the conformal frame. The Lyapunov exponent was used to study phase transitions based on characteristic cusp-like structures, while exhibiting different behaviors and magnitudes between the Einstein and string frames. These frame-dependent features observed in the Lyapunov exponent are determined by the balance among the dilaton field, electromagnetic force, and centrifugal effects. In contrast to coarse-grained thermodynamic quantities, which remain independent of the choice of frame under proper definitions, the Lyapunov exponent provides a fine-grained dynamical probe that distinguishes frame-dependent effects arising from the local dilaton coupling to the probe particle. Thus, orbital dynamics provide a dynamical probe of black hole phase structures while being sensitive to the frame dependence of particle motion between the Einstein and string frames. In future studies, this analysis may be extended to rotating dilatonic black holes, higher-dimensional backgrounds, higher-curvature corrections, and holographic interpretations of frame-dependent dynamical observables.

\section*{Acknowledgments}
    This research was supported by Basic Science Research Program through the National Research Foundation of Korea (NRF) funded by the Ministry of Education (NRF-2022R1I1A2063176) and the Dongguk University Research Fund of 2026.

\printbibliography

\end{document}